\newif\iffull
\fulltrue

\iffull
\documentclass[]{article}
\else
\documentclass{llncs}
\fi

\usepackage[british]{babel}
\usepackage[draft]{fixme}
\usepackage{todo}
\usepackage{url}
\usepackage{hyperref}
\usepackage{breakurl}
\usepackage{xspace}
\usepackage{tabularx}
\usepackage{amsmath}
\usepackage{amssymb}
\usepackage{array}
\usepackage{color}
\usepackage{subfigure} %
\usepackage{graphicx}
\usepackage{listings}
\usepackage{multirow}
\usepackage{multicol}
\usepackage{booktabs}
\usepackage{bbm} %
\usepackage{wrapfig}
\usepackage{tikz}

\iffull
\else
\usepackage{times}
\fi

\iffull
\usepackage{authblk}
\usepackage{fullpage}
\fi

\usepackage{xspace}
\newcommand{\myfig}{Fig.~}

\newcommand{\mysec}{Sec.~}
\newcommand{\myapp}{App.~}

\newcommand{\wrt}{w.r.t.\xspace}
\newcommand{\al}{al.\xspace}
\newcommand{\eg}{e.g.\xspace}
\newcommand{\ie}{i.e.\xspace}
\newcommand{\cf}{cf.\xspace}

  \newcommand{\dfnshort}[2]{{#1}\triangleq{#2}}
  
  \newcommand{\dfn}[2]{{#1}~\triangleq~{#2}}

  \newcommand{\re}{r}

  \let\setset\setbf

  \newcommand{\evts}{\setset{E}}

  \newcommand{\lo}{\ell}

\newcommand{\stacklabel}[1]
{\stackrel{\smash{\scriptstyle\textnormal{#1}}}}
  \newcommand{\po}{\textsf{po}}
  \newcommand{\dep}{\textsf{dp}}
  \newcommand{\rf}{\textsf{rf}}
  \newcommand{\rfe}{\textsf{rfe}}
  \newcommand{\ws}{\textsf{co}}

  \newcommand{\ews}{\textsf{coe}}
  \newcommand{\fr}{\textsf{fr}}
  
  \newcommand{\efr}{\textsf{fre}}
  \newcommand{\fenced}{\textsf{fence}}

  \newcommand{\com}{\textsf{com}}

  \newcommand{\ppo}{\textsf{ppo}}

    \newcommand{\cmp}{\operatorname{\textsf{cmp}}}

\newcommand{\import}{\operatorname{import}}

\let\prog\textsf
\let\as\texttt
\let\ltest\textbf
\newcommand{\proc}[1]{\ensuremath{T_{#1}}}
\newlength{\fmtlength}

\newcommand{\sync}{\textsf{sync}}
\newcommand{\isb}{\textsf{isb}}
\newcommand{\isync}{\textsf{isync}}
\newcommand{\lwsync}{\textsf{lwsync}}
\newcommand{\mfence}{\textsf{mfence}}
\newcommand{\dmb}{\textsf{dmb}}

\newcommand{\Cands}{relations}

\newcommand{\llb}[3]{#1#2#3}
\newcommand{\safes}{\operatorname{safe}}

\newcommand{\cfg}{\textsf{cfg}}

\newcommand{\porw}{\textsf{poRW}}
\newcommand{\powr}{\textsf{poWR}}
\newcommand{\poww}{\textsf{poWW}}
\newcommand{\porr}{\textsf{poRR}}
\newcommand{\poif}{\textsf{poC}}

\newcommand{\pos}{\textsf{po}$_s$}
\newcommand{\posplus}{\textsf{po}$_s^+$}
\newcommand{\posstar}{\textsf{po}$_s^*$}
\newcommand{\ilp}{\textsf{ILP}}
\newcommand{\aeg}{\textsf{aeg}}
\newcommand{\PT}{\operatorname{\textsf{between}}}
\newcommand{\CT}{\operatorname{\textsf{cumul}}}
\newcommand{\IT}{\operatorname{\textsf{ctrl}}}
\newcommand{\cost}{\operatorname{\textsf{cost}}}

\newcommand{\APLACES}{\operatorname{\textsf{actual-places}}}
\newcommand{\PLACES}{\operatorname{\textsf{potential-places}}}

\newcolumntype{Y}{@{}r@{\,}X}
\newcommand{\instab}[2]{\ \(#2\)  & \as{#1}}

\newcommand{\pset}[2]{\(\as{#1} \leftarrow \as{#2}\)}
\newcommand{\pstore}[2]{\pset{#2}{#1}}
\newcommand{\pload}[2]{\pset{#1}{#2}}

\newcommand{\plwarx}[3]{\as{lwarx #1,#2,#3}}
\newcommand{\pstwcx}[3]{\as{stwcx. #1,#2,#3}}
\newcommand{\pstw}[3]{\as{stw #1,#2,#3}}
\newcommand{\pbne}[1]{\as{bne #1}}
\newcommand{\pcmp}[2]{\as{cmpw #1,#2}}
\newcommand{\haut}{\rule{0ex}{2ex}}
\newcommand{\bas}{\rule[-1ex]{0.5ex}{0ex}}
\newcommand{\Rmw}[1][40ex]
{\begin{tabular}{rl}
\haut \instab{\as{loop:}}{} \\
\instab{\plwarx{r1}{0}{r5}}{(a_1)} \\
\instab{[\dots]}{}\\
\instab{\pstwcx{r2}{0}{r5}}{(a_2)} \\
\bas\instab{\pbne{loop}}{(b)} \\ %
\end{tabular}}
\newcommand{\Atom}[1][40ex]
{\begin{tabular}{rl}
\haut \instab{\plwarx{r1}{0}{r5}}{(a_1)} \\
\instab{[\dots]}{}\\
\bas\instab{\pstwcx{r2}{0}{r5}}{(a_2)} \\
\end{tabular}}
\newcommand{\Lo}[1][40ex]
{\begin{tabular}{rl}
\haut\instab{\as{loop:}}{} \\
\instab{\plwarx{r6}{0}{r3}}{(a_1)} \\
\instab{\pcmp{r4}{r6}}{(b)} \\
\instab{\pbne{loop}}{(c)} \\
\instab{\pstwcx{r5}{0}{r3}}{(a_2)} \\
\instab{\pbne{loop}}{(d)} \\
\instab{\as{isync}}{(e)} \\
\bas\instab{[\dots]}{}\\ 
\end{tabular}}
\newcommand{\ULo}[1][40ex]
{\begin{tabular}{rl}
\instab{[\dots]}{} \\
\haut\instab{\as{lwsync}}{(f)} \\ 
\bas\instab{\pstw{r4}{0}{r3}}{(g)} \\
\end{tabular}}

\makeatletter
\def\url@leostyle{%
  \@ifundefined{selectfont}{\def\UrlFont{\small\sf}}{\def\UrlFont{\small\sf}}}
\makeatother
\makeatletter
\newcommand\smallish{\@setfontsize\smallish{7pt}{6}}
\makeatother

\makeatletter
\gdef\thelstlisting{\@arabic\c@lstlisting}
\makeatother

\begin{document}

\setlength{\pdfpageheight}{\paperheight}
\setlength{\pdfpagewidth}{\paperwidth}

\iffull
\title{Don't sit on the fence \\
\Large A static analysis approach to automatic fence insertion}
\else
\title{Don't sit on the fence\thanks{Supported by SRC/2269.002,
EPSRC/H017585/1 and ERC/280053.}}
\subtitle{A static analysis approach to automatic fence insertion}
\fi

\iffull
\author[1]{Jade Alglave}
\author[2]{Daniel Kroening}
\author[2]{Vincent Nimal}
\author[2]{Daniel Poetzl}
\affil[1]{\small University College London, UK}
\affil[2]{\small University of Oxford, UK}
\date{}
\else
\author{Jade Alglave\inst{1}
\and Daniel Kroening\inst{2}
\and Vincent Nimal\inst{2}
\and Daniel Poetzl\inst{2}}

\institute{\small $^{1}$University College London \quad \quad \small $^{2}$University of Oxford}
\fi

\maketitle

\begin{abstract}
Modern architectures rely on memory fences to prevent
undesired weakenings of memory consistency.  As the fences' semantics may be
subtle, the automation of their placement is highly desirable. But precise
methods for restoring consistency do not scale to deployed systems code. We choose
to trade some precision for genuine scalability: our technique is suitable for
large code bases. We implement it in our new \prog{musketeer} tool, and detail
experiments on more than 350 executables of packages found in Debian
Linux~7.1, \eg \prog{memcached} (about 10000 LoC).
\end{abstract}

\section{Introduction\label{introduction}}
\renewcommand{\figurename}{Fig.}

Concurrent programs are hard to design and implement, especially when
running on multiprocessor architectures.  Multiprocessors implement
\emph{weak memory models}, which feature \eg{} \emph{instruction
reordering}, \emph{store buffering} (both appearing on x86), or \emph{store
atomicity relaxation} (a particularity of Power and ARM).  Hence,
multiprocessors allow more behaviours than Lamport's \emph{Sequential
Consistency} (SC)~\cite{lam79}, a theoretical model where the execution of a
program corresponds to an interleaving of the different threads.  This has a
dramatic effect on programmers, most of whom learned to program with SC.

Fortunately, architectures provide special \emph{fence} (or \emph{barrier})
instructions to prevent certain behaviours.  Yet both the questions of
\emph{where} and \emph{how} to insert fences are contentious, as fences are
architecture-specific and expensive.

Attempts at automatically placing fences include Visual Studio 2013, which
offers an option to guarantee acquire/release semantics (we study the
performance impact of this policy in \mysec\ref{sec:motivation}).  The C++11
standard provides an elaborate API for inter-thread communication, giving
the programmer some control over which fences are used, and where.  But the
use of such APIs might be a hard task, even for expert programmers.  For
example, Norris and Demsky reported a bug found in a published C11
implementation of a work-stealing queue~\cite{DBLP:conf/oopsla/NorrisD13}.

We address here the question of how to \emph{synthesise} fences, \ie{}
automatically place them in a program to enforce
robustness/stability~\cite{bmm11,am11} (which implies SC).  This should
lighten the programmer's burden.  The fence synthesis tool needs to be based
on a precise model of weak memory.  In verification, models commonly adopt
an \emph{operational} style, where an execution is an interleaving of
transitions accessing the memory (as in SC).  To address weaker architectures,
the models are augmented with buffers and queues that implement the features
of the hardware.  Similarly, a good fraction of the fence synthesis methods,
\eg~\cite{DBLP:conf/tacas/LindenW13,DBLP:conf/fmcad/KupersteinVY10,DBLP:conf/pldi/KupersteinVY11,DBLP:conf/pldi/LiuNPVY12,DBLP:conf/tacas/AbdullaACLR13,DBLP:conf/esop/BouajjaniDM13}
(see also \myfig\ref{table}), rely on operational models to describe
executions of programs.

\paragraph{Challenges} Thus, methods using operational models
inherit the limitations of methods based on interleavings, \emph{e.g.} the
\emph{``severely limited scalability''}, as~\cite{DBLP:conf/pldi/LiuNPVY12}
puts it.  Indeed, none of them scale to programs with more than a few
hundred lines of code, due to the very large number of executions a program can
have.  Another impediment to scalability is that these methods establish if
there is a need for fences by exploring the executions of a program one by one.

Finally, considering models \`a la Power makes the problem significantly more
difficult. Intel x86 offers only one fence (\mfence{}), but Power offers a
variety of synchronisation: fences (\eg \sync{} and \lwsync{}), or dependencies
(address, data or control). This diversity makes the optimisation more subtle:
one cannot simply minimise the number of fences, but rather has to consider the
costs of the different synchronisation mechanisms; it might be cheaper to use
one full fence than four dependencies.

\paragraph{Our approach} We tackle these challenges with a static approach. 
Our choice of model almost mandates this approach: we rely on the axiomatic
semantics of~\cite{ams10}.  We feel that an axiomatic semantics is an
invitation to build abstract objects that embrace all the executions of a
program.

Previous works, \eg~\cite{ss88,am11,bmm11,DBLP:conf/esop/BouajjaniDM13}, show
that weak memory behaviours boil down to the presence of certain cycles, called
\emph{critical cycles}, in the executions of the program. A critical cycle
essentially represents a minimal violation of SC, and thus indicates where to place
fences to restore SC. We detect these cycles statically, by exploring an
over-approximation of the executions of the program.  

\paragraph{Contributions} Our method is sound for a wide range of
architectures, including x86-TSO, Power and ARM; and scales for large code
bases, such as \prog{memcached} (about $10 000$ LoC). We implemented it in our
new \prog{musketeer} tool. Our method is the most precise of the static
analysis methods (see \mysec\ref{sec:motivation}). To do this comparison, we
implemented all these methods in our tool; for example, the \prog{pensieve}
policy~\cite{DBLP:conf/ppopp/SuraFWMLP05} was designed for Java only, and we now provide it for
x86-TSO, Power and ARM. Thus, our tool \prog{musketeer} gives a comparison point
for the field.

\paragraph{Outline}
We discuss the performance impact of fences in \mysec\ref{sec:motivation}, and
survey related work in~\mysec\ref{related-work}. We recall our weak memory semantics
in \mysec\ref{context}. We detail how we detect critical cycles in
\mysec\ref{detection}, and how we place fences in \mysec\ref{inference}.
In \mysec\ref{sec:comparison}, we compare existing tools and our new tool
\prog{musketeer}. 
We provide the sources, benchmarks and experimental reports
online at \url{http://www.cprover.org/wmm/musketeer}.  

\vspace*{-3mm}
\section{Motivation}\label{sec:motivation}

\setlength{\parskip}{0pt}

Before optimising the placement of fences, we investigated whether naive
approaches to fence insertion indeed have a negative performance impact.
To that end, we measured the
overhead of different fencing methods on a stack and a queue from the
\textsf{liblfds} lock-free data structure package (\url{http://liblfds.org}).
For each data structure, we built a harness (consisting of 4 threads) that
concurrently invokes its operations.
We built several versions of the above two programs:
\vspace*{-2mm}
\begin{itemize}
\item \textsc{(m)} with fences inserted by our tool \prog{musketeer};
\item \textsc{(p)} with fences following the \emph{delay set analysis} of the
\prog{pensieve} compiler~\cite{DBLP:conf/ppopp/SuraFWMLP05}, \ie a static
over-approximation of Shasha and Snir's eponymous (dynamic)
analysis~\cite{ss88} (see also the discussion of Lee and Padua's
work~\cite{lp01} in \mysec\ref{related-work});
\item \textsc{(v)} with fences following the \emph{Visual Studio} policy,
\ie guaranteeing acquire/release semantics (in the C11 sense~\cite{c11}), but not
SC, for
reads and writes of \texttt{volatile} variables (see
\url{http://msdn.microsoft.com/en-us/library/vstudio/jj635841.aspx}, accessed 04-11-2013). On x86, no fences are necessary as the model is
sufficiently strong already; hence, we only provide data for ARM;
\item \textsc{(e)} with fences %
after each
access to a shared variable;
\item \textsc{(h)} with an \mfence{} (x86) or a \dmb{} (ARM) after
every assembly instruction that writes (x86) or reads or writes (ARM) \emph{static global} or \emph{heap
data}.
\end{itemize}

\vspace*{-1mm}

We emphasise that these experiments required us to implement \textsc{(P)},
\textsc{(E)} and \textsc{(V)} ourselves, so that they would handle the
architectures that we considered. This means in particular that our tool
provides the \prog{pensieve} policy \textsc{(P)} for TSO, Power and ARM,
whereas the original \prog{pensieve} targeted Java only.

\iffull
\begin{figure}[]%
\centering
\subfigure{
\includegraphics[scale=0.75]{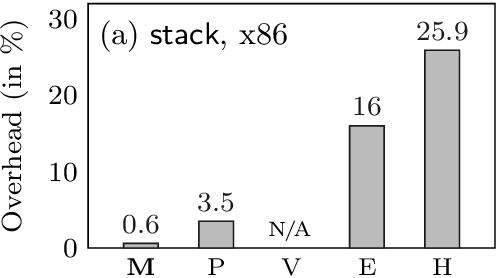}
}
\subfigure{
\includegraphics[scale=0.75]{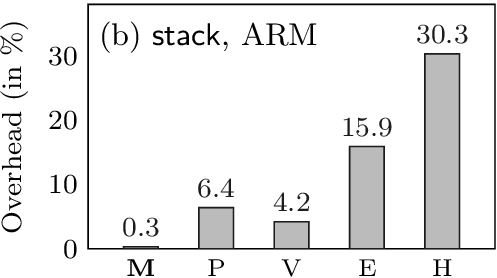}
}
\subfigure{
\includegraphics[scale=0.75]{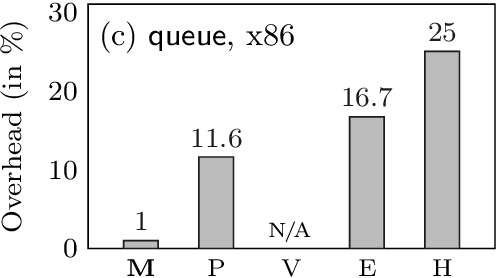}
}
\subfigure{
\includegraphics[scale=0.75]{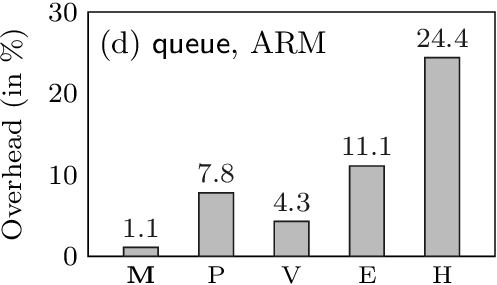}
}
\caption{Overheads for the different fencing strategies}
\label{fig:performance-x86}
\end{figure}%
\else
\begin{figure}[!t]%
\centering
\subfigure{
\includegraphics[scale=0.85]{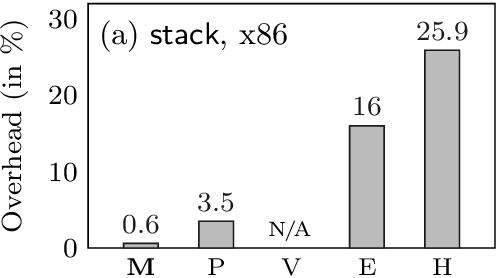}
}
\subfigure{
\includegraphics[scale=0.85]{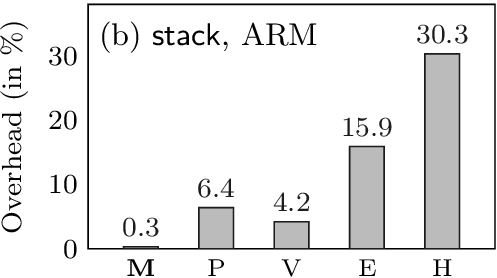}
}
\subfigure{
\includegraphics[scale=0.85]{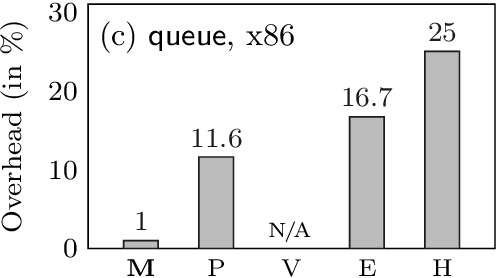}
}
\subfigure{
\includegraphics[scale=0.85]{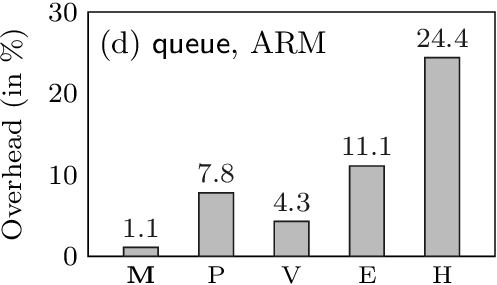}
}
\caption{Overheads for the different fencing strategies}
\label{fig:performance-x86}
\end{figure}%
\fi

\noindent
We ran all versions $100$ times, on an x86-64 Intel Core i5-3570 with 4 cores
(3.40\,GHz) and 4\,GB of RAM, and on an ARMv7 (32-bit) Samsung Exynos 4412 with
4 cores (1.6\,GHz) and 2\,GB of RAM.

For each program version, \myfig\ref{fig:performance-x86} shows the mean
overhead \wrt{} the unfenced program. We give the overhead \iffull(in \%) \fi
 in \emph{user time} (as given by Linux \texttt{time}),   
\ie the time spent by the program in user mode on the CPU.  
\iffull
\else
We refer the
reader to our study of the statistical significance of these experiments
(using confidence intervals)
in the full version of this paper~\cite{DBLP:journals/corr/AlglaveKNP13}.
\fi
Amongst the approaches that guarantee SC (\ie all but~\textsc{v}), the best
results were achieved with our tool \prog{musketeer}.

\iffull
We checked the statistical significance of the execution time improvement of 
our method over the existing methods by computing and comparing the confidence 
intervals for a sample size of $N=100$ and a confidence level $1-\alpha=95\%$
in \myfig\ref{fig:ds-exp}. If the confidence intervals for two methods are 
non-overlapping, we can conclude that the difference between the means is 
statistically significant.

\begin{figure}[b]
\centering
\begin{tabular}{l|c|c|c|c}
& \prog{stack} on x86 & \prog{stack} on ARM & \prog{queue} on x86 & \prog{queue} on ARM\\
\hline
\textsc{(o)} & [9.757; 9.798] & [11.291; 11.369] & [11.947; 11.978] & [20.441; 20.634]\\
\textsc{(m)} & [9.818; 9.850] & [11.316; 11.408] & [12.067; 12.099] & [20.687; 20.857]\\
\textsc{(p)} & [10.077; 10.155] & [11.995; 12.109] & [13.339; 13.373] & [22.035; 22.240]\\
\textsc{(v)} &        N/A       & [11.779; 11.834] &       N/A        & [21.334; 21.526]\\
\textsc{(e)} & [11.316; 11.360] & [13.071; 13.200] & [13.949, 13.981] & [22.722; 22.903]\\
\textsc{(h)} & [12.286; 12.325] & [14.676; 14.844] & [14.941, 14.963] & [25.468; 25.633]
\end{tabular}
\caption{Confidence intervals for data structure experiments.}
\label{fig:ds-exp}
\end{figure}
\fi
\section{Related work}\label{related-work}

\iffull
\begin{figure}[]
\else
\begin{wrapfigure}[]{R}{0.57\linewidth}
\fi
\centering
\iffull
\else
\vspace*{-0.8cm}
\fi
\scalebox{\iffull1\else0.8\fi}{
\begin{tabular}{c|c|c|c} 
authors & tool %
& model style & objective\\
\hline
Abdulla et \al{} \cite{DBLP:conf/tacas/AbdullaACLR13}
& \prog{memorax} 
& operational & reachability
\\
Alglave et \al{} \cite{ams10}
& \prog{offence} & 
axiomatic & SC
\\
Bouajjani et \al{} \cite{DBLP:conf/esop/BouajjaniDM13}
& \prog{trencher} 
& operational & SC
\\
Fang et \al{} \cite{DBLP:conf/ics/FangLM03}
& \prog{pensieve}
& axiomatic & SC
\\
Kuperstein et \al{} \cite{DBLP:conf/fmcad/KupersteinVY10}
& \prog{fender} 
& operational & reachability
\\
Kuperstein et \al{} \cite{DBLP:conf/pldi/KupersteinVY11}
& \prog{blender} 
& operational & reachability
\\
Linden et \al{} \cite{DBLP:conf/tacas/LindenW13}
& \prog{remmex} 
& operational & reachability
\\
Liu et \al{} \cite{DBLP:conf/pldi/LiuNPVY12}
& \prog{dfence} 
& operational & specification
\\
Sura et \al{} \cite{DBLP:conf/ppopp/SuraFWMLP05}
& \prog{pensieve}
& axiomatic & SC
\end{tabular}
}
\caption{Fence synthesis tools}
\label{table}
\iffull
\end{figure}
\else
\vspace{-0.3cm}
\end{wrapfigure}
\fi

The work of Shasha and Snir~\cite{ss88} is a foundation for the field  of fence
synthesis.  Most of the work cited below inherits their notions of \emph{delay}
and \emph{critical cycle}. A delay is a pair of instructions in a thread that
can be reordered by the underlying architecture.  A critical cycle essentially
represents a minimal violation of SC.  \myfig\ref{table} classifies the methods
mentioned in this section \wrt their style of model (operational or axiomatic).
We report our experimental comparison of these tools in
\mysec\ref{sec:comparison}.  Below, we detail fence synthesis methods per
style. We write TSO for Total Store Order, implemented in Sparc
TSO~\cite{sparc:94} and Intel x86~\cite{oss09}.  We write PSO for Partial Store
Order and RMO for Relaxed Memory Order, two other Sparc architectures. We write
Power for IBM Power~\cite{ppc:2.06}.

\paragraph{Operational models}

Linden and Wolper~\cite{DBLP:conf/tacas/LindenW13} explore all executions
(using what they call \emph{automata acceleration}) to simulate the reorderings
occuring under TSO and PSO.
Abdulla et \al{}~\cite{DBLP:conf/tacas/AbdullaACLR13} couple predicate
abstraction for TSO with a counterexample-guided strategy.
They check if an error state is reachable; if so, they calculate what they
call the \emph{maximal permissive} sets of fences that forbid this error
state.  Their method guarantees that the fences they find are
\emph{necessary}, i.e., removing a fence from the set would make
the error state reachable again.

Kuperstein et \al{}~\cite{DBLP:conf/fmcad/KupersteinVY10} explore all
executions for TSO, PSO and a subset of RMO, and along the way build
constraints encoding reorderings leading to error states. The fences can be
derived from the set of constraints at the error states. The same
authors~\cite{DBLP:conf/pldi/KupersteinVY11} improve this exploration under TSO
and PSO using an abstract interpretation they call \emph{partial coherence
abstraction}, relaxing the order in the write buffers after a certain
bound, thus reducing the state space to explore. 
Liu et \al{}~\cite{DBLP:conf/pldi/LiuNPVY12} offer a \emph{dynamic synthesis}
approach for TSO and PSO, enumerating the possible sets of fences to prevent
an execution picked dynamically from reaching an error state.

Bouajjani et \al{}~\cite{DBLP:conf/esop/BouajjaniDM13} build on an operational
model of TSO.
They
look for \emph{minimum violations} (viz. critical cycles) by enumerating
\emph{attackers} (viz. delays). Like us, they use linear programming.  However,
they first enumerate all the solutions, then encode them as an \ilp{}, and
finally ask the solver to pick the least expensive one. Our method directly
encodes the whole decision problem as an \ilp{}. The solver thus both constructs the
solution (avoiding the exponential-size \ilp{} problem) and ensures its
optimality. 

All the approaches above focus on TSO and its siblings PSO
and RMO, whereas we also handle the significantly weaker Power, including
quite subtle barriers (\eg \lwsync{}) compared to the simpler
\mfence{} of x86.

\paragraph{Axiomatic models}

Krishnamurthy et \al{}~\cite{DBLP:journals/jpdc/KrishnamurthyY96} apply Shasha
and Snir's method to \emph{single program multiple data} systems.  Their
abstraction is similar to ours, except that they do not handle
pointers. %

Lee and Padua~\cite{lp01} propose an algorithm based on Shasha and Snir's work.
They use dominators in graphs to determine which fences are redundant. This
approach was later implemented by Fang et al.~\cite{DBLP:conf/ics/FangLM03} in
\prog{pensieve}, a compiler for Java. 
Sura et \al{} later implemented a more precise approach in
\prog{pensieve}~\cite{DBLP:conf/ppopp/SuraFWMLP05} (see \textsc{(p)} in
\mysec\ref{sec:motivation}). They pair the cycle detection with an analysis to
detect synchronisation that could prevent cycles. 

Alglave and Maranget~\cite{ams10} revisit Shasha and Snir for contemporary
memory models and insert fences following a refinement of~\cite{lp01}.
Their \prog{offence} tool handles snippets of assembly code only, where the
memory locations need to be explicitly given.
\paragraph{Others}

We cite the work of Vafeiadis and Zappa
Nardelli~\cite{DBLP:conf/sas/VafeiadisN11}, who present an optimisation of
the certified \prog{CompCert-TSO} compiler to remove redundant fences on
TSO.  Marino et \al{}~\cite{DBLP:conf/pldi/MarinoSMMN11} experiment with an
SC-preserving compiler, showing overheads of no more than $34\,\%$. 
Nevertheless, they emphasise that \emph{``the overheads, however small,
might be unacceptable for certain applications''}.

\section{Axiomatic memory model\label{context}}

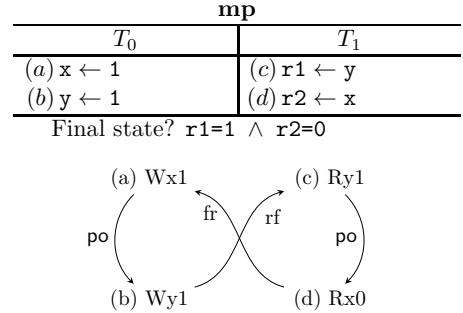
\begin{wrapfigure}[]{R}{0.4\linewidth}
\iffull\else\vspace{-1.cm}\fi
\begin{tabular}{m{\linewidth}}
\centerline{
\scalebox{0.9}{\begin{tabularx}{\linewidth}{Y|Y}
\multicolumn{4}{c}{\ltest{mp}} \\ \hline
\multicolumn{2}{c|}{\haut\proc{0}} &
\multicolumn{2}{c}{\haut\proc{1}} \\ \hline
\haut\instab{\pstore{1}{x}}{(a)} & \instab{\pload{r1}{y}}{(c)} \\
\bas\instab{\pstore{1}{y}}{(b)} & \instab{\pload{r2}{x}}{(d)} \\ \hline
\multicolumn{4}{l}{~~~Final state? \as{r1=1 $\wedge$ \as{r2=0}}}
\end{tabularx}
}}
\\
\centerline{
  \scalebox{0.8}{
  \begin{tikzpicture}[>=stealth,thin,inner sep=0pt,text centered,shape=rectangle]
  \useasboundingbox (1cm,-1.8cm) rectangle (1.9cm,0cm);
    \begin{scope}[minimum height=0.5cm,minimum width=0.5cm,text width=1.5cm]
      \node (a)  at (0, 0)  {{(a) Wx1}};
      \node (b)  at (0, -2)  {(b) Wy1};
      \node (c)  at (3, 0)  {(c) Ry1};
      \node (d)  at (3, -2)  {(d) Rx0};
    \end{scope}
    \path[->] (a) edge [out=225,in=135] node [left=0.1cm] {\po} (b);
    \path[->] (b) edge [out=15,in=195]  node [pos=0.8,below right=0.07cm] {rf} (c);
    \path[->] (c) edge [out=-45,in=45]  node [left=0.1cm] {\po} (d);
    \path[->] (d) edge [out=165,in=-15] node [pos=0.8,below left==0.07cm] {fr} (a);
  \end{tikzpicture}
}
}
\end{tabular}
\caption{\label{fig:mp} Message Passing (\ltest{mp})}
\vspace{-0.3cm}
\end{wrapfigure}

Weak memory can occur as follows: a thread sends a write to a store buffer,
then a cache, and finally to memory.  %
While the write transits through buffers and caches, a read can occur
before the value is available to all threads in memory.

To describe such situations, we use the framework
of~\cite{ams10}, embracing in particular SC, Sun TSO (\ie
the x86 model~\cite{oss09}), and a fragment of Power. The core of this
framework consists of \emph{relations} over \emph{memory events}.  

We illustrate this framework using a \emph{litmus test} (\myfig\ref{fig:mp}).
The top shows a multi-threaded program. The shared variables {\tt
x} and {\tt y} are assumed to be initialised to zero. 
A store instruction (\eg \pstore{1}{x}
on $T_0$) gives rise to a write event (\llb{$(a)$}{W}{x}{$1$}), and a load
instruction (\eg \pload{r1}{y} on $T_1$) to a read event
(\llb{$(c)$}{R}{y}{1}).
The bottom of \myfig\ref{fig:mp}
shows one particular execution of the program (also called \emph{event graph}),
corresponding to the final state {\tt r1=1} and {\tt r2=0}.

In the framework of~\cite{ams10}, an execution that is not
possible on SC has a cyclic event graph (as the one shown in \myfig\ref{fig:mp}).
A weaker architecture may \emph{relax} some of the relations contributing to a
cycle. If the removal of the relaxed edges from the event graph makes it
acyclic, the architecture allows the execution. For example, Power relaxes
the program order \po{} (amongst other things), thereby making the graph in
\myfig\ref{fig:mp} acyclic. Hence, the given execution is allowed on Power.

\paragraph{Formalisation}

An \emph{event} is a memory read or a write to memory, composed of a
unique identifier, a direction (R for read or W for write), a memory address,
and a value. We represent each instruction by the events it issues.
In~\myfig\ref{fig:mp}, we associate the store instruction \pstore{1}{$x$} in thread 
$T_0$ with the event~\llb{$(a)$}{W}{$x$}{$1$}.   

A set of events~$\evts$ and their program order~$\po$ form an \emph{event
structure} $\dfnshort{E}{(\evts,\po)}$. The program order $\po$~is a per-thread
total order over $\evts$.  We write $\dep$ (with $\dep \subseteq
\po$) for the relation that models \emph{dependencies} between
instructions. For instance, there is a \emph{data dependency} between a load and
a store when the value written by the store was computed from the value
obtained by the load.

We represent the \emph{communication} between threads via an \emph{execution
witness} $\dfnshort{X}{(\ws,\rf)}$, which consists of two relations over the
events.  First, the \emph{coherence} $\ws$ is a per-address total order on
write events which models the \emph{memory coherence} widely assumed by modern
architectures%
. It links a write $w$ to any write $w'$ to the same address that hits the
memory after $w$.  Second, the \emph{read-from} relation $\rf$ links a
write~$w$ to a read~$\re$ such that $\re$ reads the value written by~$w$. 
Finally, we derive the \emph{from-read} relation $\fr$ from $\ws$ and $\rf$.
A~read~$r$ is in $\fr$ with a write~$w$ if the write $w'$ from which $r$
reads hits the memory before~$w$.  Formally, we have: ${(r, w) \in
\fr}\triangleq{\exists w'. (w',r) \in \rf \wedge (w',w) \in \ws}$.

In \myfig\ref{fig:mp}, the specified outcome corresponds to the execution below
if each location initially holds~$0$.  If {\tt r1=1} in the end, the
read~$(c)$ on~$T_1$ took its value from the write~$(b)$ on~$T_0$, hence~$(b,c)
\in \rf$.  If {\tt r2=0} in the end, the read~$(d)$ took its value from the
initial state, thus before the write~$(a)$ on~$T_0$, hence~$(d,a) \in \fr$. 
In the following, we write $\rfe$ (resp. $\ews,\efr$) for the
\emph{external read-from} (resp. coherence, from-read), \ie when the source and
target belong to different threads. 

\iffull
\begin{figure}[b]
\centering
\begin{tabular}{l l l l} %
& SC & x86 & Power \\
\hline
\powr{} & yes & \mfence{} & \sync{}\\
\poww{} & yes & yes & \sync{}, \lwsync{}\\
\porw{} & yes & yes & \sync{}, \lwsync{}, \dep\\
\porr{} & yes & yes & \sync, \lwsync, \dep, branch;\isync{}
\end{tabular}
\caption{$\ppo$ and fences per architecture}
\label{table_fences}
\end{figure}
\else 
\begin{wrapfigure}[]{I}{0.55\linewidth}
\vspace{-0.9cm}
\centering
{\small
\begin{tabular}{l l l l} %
& SC & x86 & Power \\
\hline
\powr{} & yes & \mfence{} & \sync{}\\
\poww{} & yes & yes & \sync{}, \lwsync{}\\
\porw{} & yes & yes & \sync{}, \lwsync{}, \dep\\
\porr{} & yes & yes & \sync, \lwsync, \dep, branch;\isync{}
\end{tabular}
}
\vspace{-1mm}
\caption{$\ppo$ and fences per architecture}
\label{table_fences}
\vspace*{-5mm}
\end{wrapfigure}
\fi

\paragraph{Relaxed or safe}

\iffull
We model the scenario of reads occurring in advance, as described at the
beginning of this section, by some subrelation of the read-from $\rf$ being
\emph{relaxed}, \ie not included in global happens before.
\fi
When a thread can
read from its own store buffer~\cite{ag96} (the typical TSO/x86 scenario), we
relax the internal read-from, that is, $\rf$ where source and target belong
to the same thread. When two threads $\proc{0}$ and $\proc{1}$ can
communicate privately via a cache (a case of \emph{write atomicity}
relaxation~\cite{ag96}), we relax the external read-from $\rfe$, and call the
corresponding write \emph{non-atomic}. This is the main particularity of Power
and ARM, and cannot happen on TSO/x86.  Some program-order pairs may be relaxed
(\eg write-read pairs on x86, and all but \dep~ones on Power), \ie only a
subset of $\po$ is guaranteed to occur in order.  This subset constitutes
the \emph{preserved program order}, $\ppo$. When a relation must not be relaxed
on a given architecture, we call it \emph{safe}.

\myfig\ref{table_fences} summarises $\ppo$ per architecture. The columns are
architectures, \eg x86, and the lines are relations, \eg \powr. We write \eg
\powr{} for the program order between a write and a read. We write ``yes'' when
the relation is in the \ppo{} of the architecture: \eg \powr{} is in the \ppo{}
of SC. When we write something else, typically the name of a fence, \eg
\mfence{}, the relation is not in the \ppo{} of the architecture (\eg \powr{}
is not in the \ppo{} of x86), and the fence can restore the ordering: \eg \mfence{}
maintains write-read pairs in program order.
Following~\cite{ams10}, the relation ${\fenced}$ (with $\fenced \subseteq
{\po}$) induced by a fence is \emph{non-cumulative} when it only orders certain
pairs of events surrounding the fence.
The relation $\fenced$ is \emph{cumulative} when it additionally makes writes
atomic, \eg by flushing caches.  In our model, this amounts to making sequences
of external read-from and fences ($\rfe;\fenced$ or $\fenced;\rfe$) safe, even
though $\rfe$ alone would not be safe. 
In \myfig\ref{fig:mp}, placing a cumulative fence between the two writes on
$T_0$ will not only prevent their reordering, but also enforce an ordering
between the write $(a)$ on $T_0$ and the read $(c)$ on $T_1$, which reads from
$T_0$.

\vspace{-1mm} 
\paragraph{Architectures}

An \emph{architecture} $A$ determines the set $\safes_A$ of \Cands~safe
on $A$.
Following~\cite{ams10}, we always consider the
coherence $\ws$, the from-read relation $\fr$ and the fences to be safe. 
SC relaxes nothing, \ie $\rf$ and $\po$ are safe.  TSO authorises the
reordering of write-read pairs and store buffering but nothing else.

\paragraph{Critical cycles} 
Following~\cite{ss88,am11}, for an architecture $A$, a \emph{delay} is a
\po\ or \rf\ edge that is not safe (\ie is relaxed) on $A$. An execution $(E,X)$ is valid on $A$ yet not
on SC iff it contains critical cycles~\cite{am11}.
Formally, a \emph{critical cycle} \wrt $A$ is a cycle
in ${\po{}} \cup {\com{}}$, where $\dfn{\com}{\ws \cup \rf \cup \fr}$ is the
\emph{communication relation}, which has the following characteristics (the
last two ensure the minimality of the critical cycles):
(1) the cycle contains at least one delay for $A$;
(2) per thread, (i) there are at most two accesses $a$ and $b$,
and (ii) they access distinct memory locations; and
(3) for a memory location $\lo$, there are at most three accesses to
$\lo$ along the cycle, which belong to distinct threads.  

\myfig\ref{fig:mp} shows a critical cycle \wrt Power. The \po{} edge on $T_0$,
the \po{} edge on $T_1$, and the \rf{} edge between $T_0$ and $T_1$, are all
unsafe on Power.
On the other hand, the cycle
in \myfig\ref{fig:mp} does not contain a delay \wrt TSO, and is thus not a
critical cycle on TSO.

To forbid executions containing critical cycles, one can insert fences into
the program to prevent delays. To prevent a \po{} delay, a fence can
be inserted between the two accesses forming the delay,
following~\myfig\ref{table_fences}. To
prevent an \rf{} delay, a cumulative fence must be used (see \mysec\ref{inference}
for details).
For the example in \myfig\ref{fig:mp}, for Power, we need to place a cumulative
fence between the two writes on $T_0$, preventing both the \po{} and the
adjacent \rf{} edge from being relaxed, and use a dependency or fence to prevent
the \po{} edge on $T_1$ from being relaxed.
\section{Static detection of critical cycles}\label{detection}

We want to synthesise fences to prevent weak behaviours and thus restore SC.
We explained in \mysec\ref{context} that we should place fences along the
critical cycles of the program executions.  To find the critical cycles, we
look for cycles in an over-approximation of all the executions of the program.
We hence avoid enumeration of all traces, which would hinder scalability, and
get all the critical cycles of all program executions at once.  Thus we can
find all fences preventing the critical cycles corresponding to two executions
in one step, instead of examining the two executions separately.

\lstset{numbers=none}

\iffull
\begin{figure}
\centering
\begin{tabular}{l|@{\quad}l}
\begin{minipage}{0.5\textwidth}
\begin{lstlisting}[multicols=2,basicstyle=\small\sf]
void thread_1(int input)
{
  int r1;
  x = input;
  if(rand()%2)
    y = 1;
  else
    r1 = z;
  x = 1;
}
void thread_2()
{
  int r2, r3, r4;
  r2 = y;
  r3 = z;
  r4 = x;
}
(*@\phantom{.}@*)
(*@\phantom{.}@*)
(*@\phantom{.}@*)
\end{lstlisting}
\end{minipage}
&
\begin{minipage}{0.4\textwidth}
\begin{lstlisting}[multicols=2, basicstyle=\small\sf]
thread_1 
   int r1;
   x = input;
   _Bool tmp;
   tmp = rand();
   [!tmp%2] goto 1;
   y = 1;
   goto 2;
1: r1 = z;
2: x = 1;
   end_function
 thread_2 
    int r2, r3, r4;
    r2 = y;
    r3 = z;
    r4 = x;
    end_function
(*@\phantom{.}@*)
(*@\phantom{.}@*)
(*@\phantom{.}@*)
(*@\phantom{.}@*)
(*@\phantom{.}@*)
\end{lstlisting}
\end{minipage}
\end{tabular}
\caption{A C program (left) and its goto-program (right)\label{AEgraph}}
\end{figure}
\else
\begin{figure}
\centering
\begin{tabular}{l|@{\quad}l}
\begin{minipage}{0.5\textwidth}
\begin{lstlisting}[multicols=2,basicstyle=\smallish\sf]
void thread_1(int input)
{
  int r1;
  x = input;
  if(rand()%
    y = 1;
  else
    r1 = z;
  x = 1;
}
void thread_2()
{
  int r2, r3, r4;
  r2 = y;
  r3 = z;
  r4 = x;
}
(*@\phantom{.}@*)
(*@\phantom{.}@*)
(*@\phantom{.}@*)
(*@\phantom{.}@*)
\end{lstlisting}
\end{minipage}
&
\begin{minipage}{0.4\textwidth}
\begin{lstlisting}[multicols=2, basicstyle=\smallish\sf]
thread_1 
   int r1;
   x = input;
   _Bool tmp;
   tmp = rand();
   [!tmp%
   y = 1;
   goto 2;
1: r1 = z;
2: x = 1;
   end_function
 thread_2 
    int r2, r3, r4;
    r2 = y;
    r3 = z;
    r4 = x;
    end_function
(*@\phantom{.}@*)
(*@\phantom{.}@*)
(*@\phantom{.}@*)
(*@\phantom{.}@*)
(*@\phantom{.}@*)
(*@\phantom{.}@*)
\end{lstlisting}
\end{minipage}
\end{tabular}
\caption{A C program (left) and its goto-program (right)\label{AEgraph}}
\end{figure}
\fi

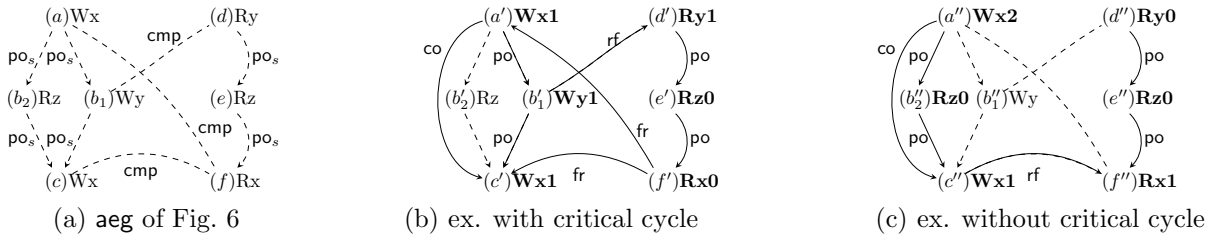
\begin{figure}[b]
\centering
\begin{tabular}{p{0.30\linewidth} p{0.34\linewidth} p{0.33\linewidth}}%
  \scalebox{0.73}{
  \begin{tikzpicture}[>=stealth,thin,inner sep=0pt,text centered,shape=rectangle]

    \begin{scope}[minimum height=0.5cm,minimum width=0.5cm,text width=0.5cm]
      \node (1)  at (-1, 0)  {($a$)Wx};
      \node (2)  at (-0.3, -1.5)  {($b_1$)Wy};
      \node (3)  at (-1, -3)  {($c$)Wx};
      \node (4)  at (2, 0)  {($d$)Ry};
      \node (5)  at (2, -1.5)  {($e$)Rz};
      \node (6)  at (2, -3)  {($f$)Rx};
      \node (7)  at (-1.7, -1.5) {($b_2$)Rz};
    \end{scope}

\path[->] (1) edge [dashed]%
node[left=0.5mm] {\pos} (2);
\path[->] (1) edge [dashed]%
node[left=0.5mm] {\pos} (7);
\path[->] (2) edge [dashed]%
node[left=0.5mm] {\pos} (3);
\path[->] (7) edge [dashed]%
node[left=0.5mm] {\pos} (3);
\path[->] (4) edge [out=-45,in=45,dashed] node[right=0.5mm] {\pos} (5);
\path[->] (5) edge [out=-45,in=45,dashed] node[right=0.5mm] {\pos} (6);

\path[-] (1) edge [out=-45+15,in=135-15,dashed] node[pos=0.8,right=1.1mm]
  {$\cmp$} (6);
\path[-] (2) edge [out=30,in=210,dashed] node[pos=0.8,left=1.1mm] {$\cmp$} (4);
\path[-] (3) edge [bend left, dashed] node[pos=0.5,below=2.5mm] {$\cmp$} (6);
\end{tikzpicture}}
  &
  \scalebox{0.73}{ 
  \begin{tikzpicture}[>=stealth,thin,inner sep=0pt,text centered,shape=rectangle] 

    \begin{scope}[minimum height=0.5cm,minimum width=0.5cm,text width=0.5cm] 
      \node (1)  at (-1, 0)  {($a'$)\textbf{Wx1}};
      \node (2)  at (-0.3, -1.5)  {($b'_1$)\textbf{Wy1}};
      \node (3)  at (-1, -3)  {($c'$)\textbf{Wx1}};
      \node (4)  at (2, 0)  {($d'$)\textbf{Ry1}};
      \node (5)  at (2, -1.5)  {($e'$)\textbf{Rz0}};
      \node (6)  at (2, -3)  {($f'$)\textbf{Rx0}};
      \node (7)  at (-1.7, -1.5) {($b'_2$)Rz};
    \end{scope} 

\path[->] (1) edge [dashed]%
node[left=0.5mm] {\po} (2);
\path[->] (1) edge [dashed]%
node[left=0.5mm] {} (7); 
\path[->] (2) edge [dashed]%
node[left=0.5mm] {\po} (3); 
\path[->] (7) edge [dashed]%
node[left=0.5mm] {} (3); 
\path[->] (4) edge [out=-45,in=45] node[right=0.5mm] {\po} (5); 
\path[->] (5) edge [out=-45,in=45] node[right=0.5mm] {\po} (6); 

\path[<-] (1) edge [out=-45+15,in=135-15] node[pos=0.8,right=1.1mm]  
  {$\fr$} (6);

\path[-] (2) edge [out=30,in=210,dashed] node[pos=0.8,left=1.1mm] {$\rf$} (4); 

\path[<-] (3) edge [bend left] node[pos=0.5,below=2.5mm] {\fr} (6);

\path[->] (1) edge [] (2);
\path[->] (2) edge [] (3);
\path[->] (2) edge [out=30,in=210] (4);

\path[->] (1) edge [out=-45-120,in=45+120] node [color=black,
  left=-1mm,pos=0.25, text width=0.8cm] {\ws} (3);

\end{tikzpicture}} 
  &
  \scalebox{0.73}{ 
  \begin{tikzpicture}[>=stealth,thin,inner sep=0pt,text centered,shape=rectangle] 

    \begin{scope}[minimum height=0.5cm,minimum width=0.5cm,text width=0.5cm] 
      \node (1)  at (-1, 0)  {($a''$)\textbf{Wx2}};
      \node (2)  at (-0.3, -1.5)  {($b''_1$)Wy};
      \node (3)  at (-1, -3)  {($c''$)\textbf{Wx1}};
      \node (4)  at (2, 0)  {($d''$)\textbf{Ry0}};
      \node (5)  at (2, -1.5)  {($e''$)\textbf{Rz0}};
      \node (6)  at (2, -3)  {($f''$)\textbf{Rx1}};
      \node (7)  at (-1.7, -1.5) {($b''_2$)\textbf{Rz0}};
    \end{scope} 

\path[->] (1) edge [dashed]%
node[left=0.5mm] {} (2); 
\path[->] (1) edge []%
node[left=0.5mm] {\po} (7); 
\path[->] (2) edge [dashed]%
node[left=0.5mm] {} (3); 
\path[->] (7) edge []%
node[left=0.5mm] {\po} (3); 
\path[->] (4) edge [out=-45,in=45] node[right=0.5mm] {\po} (5); 
\path[->] (5) edge [out=-45,in=45] node[right=0.5mm] {\po} (6); 

\path[-] (1) edge [out=-45+15,in=135-15,dashed] node[pos=0.8,right=1.1mm]  
  {} (6); 
\path[-] (2) edge [out=30,in=210,dashed] node[pos=0.8,left=1.1mm] {} (4); 
\path[-] (3) edge [bend left, dashed] node[pos=0.5,below=2.5mm] {$\rf$} (6);

\path[->] (3) edge [bend left] (6);

\path[->] (1) edge [out=-45-120,in=45+120] node [color=black,
  left=-1mm,pos=0.25, text width=0.8cm] {\ws} (3);

\end{tikzpicture}} 
\\
~~~~~ (a) \aeg{} of \myfig\ref{AEgraph} & 
(b) ex. with critical cycle & 
~ (c) ex. without critical cycle %
\end{tabular}
\caption{The \aeg{} of \myfig\ref{AEgraph} and two executions corresponding to 
it}\label{fig:aeg-dyn}
\end{figure}

To analyse a C program, \eg on the left-hand side of \myfig\ref{AEgraph}, 
we convert it to a \emph{goto-program} (right-hand side of 
\myfig\ref{AEgraph}), the internal
representation of the CProver framework; we refer to
\url{http://www.cprover.org/goto-cc} for details.  The pointer analysis we
use is a standard concurrent points-to analysis that we have shown to be
sound for our weak memory models in earlier work~\cite{alg11}.  
\iffull
We explain in details how we handle pointers at the end of this section.
\else
A full explanation of how we handle
pointers is available in~\cite{DBLP:journals/corr/AlglaveKNP13}.
\fi
The C program
in \myfig\ref{AEgraph} features two threads which can interfere.  The first
thread writes the argument ``input'' to $x$, then randomly writes $1$ to $y$
or reads $z$, and then writes $1$ to $x$.  The second thread successively
reads $y$, $z$ and $x$.
In the corresponding goto-program, the {\bf
if-else} structure has been transformed into a guard with the condition of the
{\bf if} followed by a goto construct.
From the goto-program, we then compute an \emph{abstract event graph}
(\aeg), shown in \myfig\ref{fig:aeg-dyn}(a).  The events $a, b_1, b_2$ and
$c$ (resp.  $d, e$ and $f$) correspond to thread$_1$ (resp.  thread$_2$) in
\myfig\ref{AEgraph}.  We only consider accesses to shared variables, and
ignore the local variables.  We finally explore the \aeg{} to find the
potential critical cycles.

An \aeg{} represents all the executions of a program (in the sense of
\mysec\ref{context}).  \myfig\ref{fig:aeg-dyn}(b) and (c) give two executions
associated with the \aeg{} shown in \myfig\ref{fig:aeg-dyn}(a). For readability,
the transitive \po\ edges have been omitted (\eg between the two events $d'$ and
$f'$). The concrete events that occur in an execution are shown in bold.
In an \aeg{}, the events do not have concrete values, whereas in an
execution they do.  Also, an \aeg{} merely indicates that two accesses to the
same variable could form a data race (see the competing pairs ($\cmp$)
relation in \myfig\ref{fig:aeg-dyn}(a), which is a symmetric relation),
whereas an execution has oriented relations (\eg indicating the write that a
read takes its value from, see \eg{} the $\rf$ arrow in
\myfig\ref{fig:aeg-dyn}(b) and (c)).  The execution in
\myfig\ref{fig:aeg-dyn}(b) has a critical cycle (with respect to \eg Power)
between the events $a'$, $b_2'$, $d'$, and $f'$.  The execution in
\myfig\ref{fig:aeg-dyn}(c) does not have a critical cycle.

We build an \aeg{} essentially as in~\cite{DBLP:conf/esop/AlglaveKNT13}.
However, our goal and theirs are not the same: they instrument an input program
to reuse SC verification tools to perform weak memory verification, whereas we
are interested in automatic fence placement. Moreover, the work
of~\cite{DBLP:conf/esop/AlglaveKNT13} did not present a semantics of
goto-programs in terms of \aeg{}s, which we do in this section.

\iffull
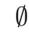
\begin{figure}[]
\centering
\begin{tabular}{%
m{0.6\linewidth} c} %
\multicolumn{2}{l}{\textsf{\large (1) assignment:}\quad $lhs$=$rhs$; $i$} \\[1mm]
$\tau[lhs=rhs; i](aeg) = \tau[i]((
aeg.\mathbb{E}_s\cup\text{evts}(lhs)\cup\text{evts}(rhs),
aeg.\text{\pos{}}\cup\text{end}(aeg.\text{\pos{}})\times(\text{evts}(rhs)\cup
\text{evts}(lhs)\backslash\text{trg}(lhs))\cup
(\text{evts}(rhs)\cup\text{evts}(lhs)\backslash\text{trg}(lhs))
\times\text{trg}(lhs),
aeg.\cmp
))$
&
\scalebox{0.7}{
\begin{tikzpicture}[>=stealth,thin,inner sep=0pt,shape=rectangle]
  \begin{scope}[minimum height=0.5cm,text centered, minimum width=0.5cm,text width=1cm]
    \node (0)  at (0,-1) {};
    \node (1)  at (3, -0.5) {R evts({\tt rhs})};
    \node (2)  at (3, -1.5)  {R evts({\tt lhs})$\backslash$trg({\tt lhs})};
    \node (3)  at (6, -1)  {W trg({\tt lhs})};
    \node (4)  at (8, -1) {$\tau[i]$};
  \end{scope}
  \path[->] (0) edge [dashed] node[above=0.5mm] {\pos{}} (1);
  \path[->] (0) edge [dashed] node[above=0.5mm] {\pos{}} (2);
  \path[->] (3) edge [dashed] node[above=0.5mm] {\pos{}} (4);
  \path[->] (1) edge [dashed] node[above=0.5mm] {\pos{}} (3);
  \path[->] (2) edge [dashed] node[above=0.5mm] {\pos{}} (3);
\end{tikzpicture}
}
\vspace{1mm}
\\
\hline
\multicolumn{2}{l}{}\\[1mm]
\multicolumn{2}{l}{\textsf{\large (2) function call:}\quad $fun$(); $i$}\\[1mm]
$\tau[fun(); i] = \tau[i] \circ \tau[\text{body}(fun)]$ &
\scalebox{0.7}{
\begin{tikzpicture}[>=stealth,thin,inner sep=0pt,text centered,shape=rectangle]
  \begin{scope}[minimum height=0.5cm,minimum width=0.5cm]
    \node (1)  at (0, 0) {};
    \node (3)  at (2, 0) [shape=rectangle, text width=1.6cm]
      {$\tau[\text{body}(\texttt{f})]$};
    \node (5)  at (4, 0)  {$\tau[i]$};
  \end{scope}

  \path[->] (1) edge [dashed] node[above=0.5mm] {\pos{}} (3);
  \path[->] (3) edge [dashed] node[above=0.5mm] {\pos{}} (5);
\end{tikzpicture}
}
\vspace{1mm}
\\
\hline
\multicolumn{2}{l}{}\\[1mm]
\multicolumn{2}{l}{\textsf{\large (3) guard:}\quad [$expr$] $i_1$; $i_2$}\\[1mm]
$\tau[[guard] i_1; i_2] (aeg) =$ \textbf{let} 
guarded$=\tau[i_1](aeg)$ \textbf{in}
$\tau[i_2] ((aeg.\mathbb{E}_s\cup \text{guarded}.\mathbb{E}_s,
aeg.\text{\pos{}}\cup\text{guarded}.\text{\pos{}},aeg.\cmp))$
&
\scalebox{0.7}{
\begin{tikzpicture}[>=stealth,thin,inner sep=0pt,text centered,shape=rectangle]
  \begin{scope}[minimum height=0.5cm,minimum width=0.5cm]
    \node (1)  at (0, 0) {};
    \node (2)  at (2, 0) {$\tau[i_1]$};
    \node (3)  at (4, 0) {$\tau[i_2]$};
  \end{scope}

  \path[->] (1) edge [dashed] node[above=0.5mm] {\pos{}} (2);
  \path[->] (2) edge [dashed] node[above=0.5mm] {\pos{}} (3);
  \path[->] (1) edge [bend right,dashed] %
    node[above=0.5mm] {\pos{}} (3);

\end{tikzpicture}
}
\vspace{1mm}
\\
\hline
\multicolumn{2}{l}{}\\[1mm]
\multicolumn{2}{l}{\textsf{\large (4) forward jump:} \quad goto $l$; }\\[1mm]
$\tau[goto\,\,l; i] = \tau[\text{follow}(l)]$
&
\scalebox{0.7}{
\begin{tikzpicture}[>=stealth,thin,inner sep=0pt,text centered,shape=rectangle]
  \begin{scope}[minimum height=0.5cm,minimum width=0.5cm]
    \node (1)  at (0, 0) {};
    \node (2)  at (2, 0) [shape=rectangle, text width=1.5cm]
      {$\tau[\text{follow}(l)]$};
  \end{scope}

  \path[->] (1) edge [dashed] node[above=0.5mm] {\pos{}} (2);
\end{tikzpicture}
}
\vspace{1mm}
\\
\hline
\multicolumn{2}{l}{}\\[1mm]
\multicolumn{2}{l}{\textsf{\large (5) backward jump:}\quad $l$: $i_1$; 
[$cond$] goto $l$; $i_2$}\\[1mm]
$\tau[l: i_1; [cond]\text{goto}\,\,l; i_2] (aeg) =$ \textbf{let} 
local$=\tau[i_1] (aeg)$ \textbf{in}
$\tau[i_2]((aeg.\mathbb{E}_s\cup\text{local}.\mathbb{E}_s, 
aeg.\text{\pos{}}\cup
\text{local}.\text{\pos{}} \cup \text{end}(\text{local}.\text{\pos{}})\times
\text{begin}(\text{local}.\text{\pos{}}),aeg.\cmp))$ &
\scalebox{0.7}{
\begin{tikzpicture}[>=stealth,thin,inner sep=0pt,text centered,shape=rectangle]
  \begin{scope}[minimum height=0.5cm,minimum width=0.5cm]
    \node (1)  at (0, 0) {};
    \node (2)  at (2, 0) [shape=rectangle]
      {$\tau[i_1]$};
    \node (4) at (4, 0) {$\tau[i_2]$};
  \end{scope}

  \path[->] (1) edge [dashed] node[above=0.5mm] {\pos{}} (2);
  \path[->] (2) edge [dashed] node[above=0.5mm] {\pos{}} (4);
  \path[<-] (2) edge [dashed, in=45, out=135]%
    node[above=0.5mm] {\pos{}} (2);
\end{tikzpicture}
}
\vspace{1mm}
\\
\hline
\multicolumn{2}{l}{}\\[1mm]
\multicolumn{2}{l}{\textsf{\large (6) assume / assert / skip:} 
\quad assume($\phi$); $i$ / assert($\phi$); $i$ / skip; $i$
}\\[1mm]
$\tau[assume(\phi); i] = \tau[i]$
&
\scalebox{0.7}{
\begin{tikzpicture}[>=stealth,thin,inner sep=0pt,text centered,shape=rectangle]
  \begin{scope}[minimum height=0.5cm,minimum width=0.5cm]
    \node (1)  at (0, 0) {};
    \node (2)  at (2, 0) {$\tau[i]$};
  \end{scope}

  \path[->, dashed] (1) edge node[above=0.5mm] {\pos{}} (2);
\end{tikzpicture}
}
\vspace{1mm}
\\
\hline
\multicolumn{2}{l}{}\\[1mm]
\multicolumn{2}{l}{\textsf{\large (7) atomic section:}\quad atomic\_begin; 
$i_1;$ atomic\_end; $i_2$ } \\[1mm]
$\tau[atomic\_start; i_1; atomic\_end; i_2](aeg) =$
\textbf{let} section=$\tau[i_1]((aeg.\mathbb{E}_s,aeg.\text{\pos{}}\cup
\text{end}(aeg.\text{\pos{}})\times\{\textsf{f}\}, aeg.\cmp))$ \textbf{in}
$\tau[i_2]((\text{section}.\mathbb{E}_s,
\text{section}.\text{\pos{}}\cup
\text{end}(\text{section}.\text{\pos{}})\times\{\textsf{f}\},
\text{section}.\cmp))$
&
\scalebox{0.7}{
\begin{tikzpicture}[>=stealth,thin,inner sep=0pt,text centered,shape=rectangle]
  \begin{scope}[minimum height=0.5cm,minimum width=0.5cm,text width=0.5cm]
    \node (1)  at (0, -0.5)  {};
    \node (2)  at (0, -1.5)  {};
    \node (3)  at (1.5, -1) {\textsf{f}};
    \node (4)  at (3, -1) [shape=rectangle, text width=0.75cm] {$\tau[i_1]$};
    \node (5)  at (4.5, -1)  {\textsf{f}};
    \node (6)  at (6, -1) [shape=rectangle] {$\tau[i_2]$};
  \end{scope}

  \path[->] (3) edge [dashed] node[above=0.5mm] {\pos{}} (4);
  \path[->] (4) edge [dashed] node[above=0.5mm] {\pos{}} (5);
  \path[->] (5) edge [dashed] node[above=0.5mm] {\pos{}} (6);
  \path[->] (1) edge [dashed] node[above=0.5mm] {\pos{}} (3);
  \path[->] (2) edge [dashed] node[above=0.5mm] {\pos{}} (3);
\end{tikzpicture}
}
\vspace{1mm}
\\
\hline
\multicolumn{2}{l}{}\\[1mm]
\multicolumn{2}{l}{\textsf{\large (8) new thread:}\quad start\_thread $th$; 
$i$}\\[1mm]
$\tau [start\_thread\,\,th; i](aeg) = $ \textbf{let}
local=$\tau[\text{body}(th)](\bar\emptyset)$ \textbf{and} 
main=$\tau[i](aeg)$ \textbf{and}
inter=$\tau[i](\bar\emptyset)$ \textbf{in}
$(\text{local}.\mathbb{E}_s\cup\text{main}.\mathbb{E}_s,
\text{local}.\text{\pos{}}\cup\text{main}.\text{\pos{}}, 
\text{local}.\mathbb{E}_s\otimes\text{inter}.\mathbb{E}_s)$
&
\scalebox{0.7}{
\begin{tikzpicture}[>=stealth,thin,inner sep=0pt,text centered,shape=rectangle]
  \begin{scope}[minimum height=0.5cm,minimum width=0.5cm,text width=0.5cm]
    \node (2)  at (7, -1) [shape=rectangle, text width=1.6cm] {$\tau[\text{body}(f)]$};
    \node (3)  at (2, -1) {};
    \node (4)  at (4, -1) [shape=rectangle, text width=0.5cm] {$\tau[i]$};
  \end{scope}
  \path[->] (3) edge [dashed] node[above=0.5mm] {\pos{}} (4);
  \path[-] (2) edge [dashed] node[above=0.5mm] {$\cmp$} (4);
\end{tikzpicture}
}
\vspace{1mm}
\\
\hline
\multicolumn{2}{l}{}\\[1mm]
\multicolumn{2}{l}{\textsf{\large (9) end of thread:}\quad end\_thread; }\\[1mm]
$\tau[end\_thread](aeg) = aeg$
& $\emptyset$
\vspace{1mm}
\\
\end{tabular}
\caption{Operations to create the \aeg{} of a goto-program}
\label{tr-aeg}
\end{figure}
\fi

\iffull
\paragraph{Constructing \aeg{}s\label{static}}

For each goto-program\iffull (\eg bottom of \myfig\ref{AEgraph}),\else, \fi 
we build an \aeg{}
$\triangleq ({\mathbb E}_s,\text{\pos},\cmp)$ as follows (see
\myfig\ref{fig:aeg-dyn}(a)), with the \emph{abstract events} ${\mathbb E}_s$,
the \emph{static program order} \pos{} and the \emph{competing pairs} $\cmp$.
\iffull
Given an \aeg{} $G$, we write respectively $G.{\mathbb E}_s$, 
$G.$\pos{} and $G.\cmp{}$ for the abstract events, the static program order and
the competing pairs of $G$.
\fi

\paragraph{An abstract event \!\!\!} represents all events with same program
point, memory location and direction (write or read). In
\myfig\ref{fig:aeg-dyn}(a), ($a$)Wx abstracts the events ($a'$)Wx1 and
($a''$)Wx2 in the executions of \myfig\ref{fig:aeg-dyn}(b) and (c).

\paragraph{The static program order \pos{} \!\!\!} abstracts all the
(dynamic) \po{} edges that connect two events in program order and that cannot
be decomposed as a succession of \po{} edges in this execution.  We write
\posplus{} (resp.  \posstar{}) for the transitive (resp. reflexive transitive)
closure of this relation. 

\paragraph{The competing pairs \!\!\!} $\cmp$ over-approximate the external
communications ${\ews} \cup {\rfe} \cup {\efr}$. In \myfig\ref{fig:aeg-dyn}(a),
the $\cmp$ edges $(a,f)$, $(b_1,d)$, and $(c,f)$ abstract in particular the
\efr{} edges $(f',c')$ and $(f',a')$, and the \rfe{} edge $(b_1',d')$ in
\myfig\ref{fig:aeg-dyn}(b).  We do not need to represent internal
communications as they are already covered by \posplus{}.
The \textsf{cmp} construction is similar to the first steps of static data race
detections (see \cite[Sec. 5]{DBLP:conf/sigsoft/KahlonSKZ09}), where shared
variables involved in write-read or write-write communications are collected.

\paragraph{To build the \aeg \!\!\!\!}, we define a semantics of
goto-programs in terms of abstract events, static program order and competing
pairs. 
\iffull
We detail below this semantics for each goto-statement. Each of these cases
is accompanied in \myfig\ref{tr-aeg} by a graphical representation summarising
the \aeg{} construction on the right-hand side, and a formal definition of the
semantics on the left-hand side.
\else
\myfig\ref{tr-aeg} summarises this semantics (which we detail below) for
each goto-instruction. 
\iffull
Formal definitions are provided in \myapp\ref{equations}.
\else 
We omit the formal definitions for brevity and provide them online. 
\fi
\fi
In \myfig\ref{tr-aeg}, we write $\tau[i]$ to represent the
semantics of a goto-instruction $i$. Other notations, \eg follow(f) or body(f),
are explained below.
We do not compute the values of our variables, and thus do not interpret the
expressions. In \myfig\ref{fig:aeg-dyn}(a), ($a$)Wx represents the assignment
``x = input'' in thread~1 in \myfig\ref{AEgraph} (since ``input'' is a local
variable). This abstracts the values that ``input'' could hold, \eg 1 (see
($a'$)Wx1 in \myfig\ref{fig:aeg-dyn}(b)) or 2 (see ($a''$)Wx2 in
\myfig\ref{fig:aeg-dyn}(c)). Prior to building the \aeg{}, we copy expressions
of conditions or function arguments into local variables. Thus all the work
over shared variables is handled in the assignment case. 

We now present the construction of the \aeg{} starting with the intra-thread
instructions (\eg assignments, function calls), creating \pos{} edges, then
the thread constructor, creating $\cmp{}$ edges. 

\par{\emph{Assignments} {\tt lhs}={\tt rhs}} We decompose this statement into
sets of abstract events: the reads from shared variables in {\tt rhs} and {\tt
lhs}, denoted by \iffull evts\else events\fi({\tt rhs}) and 
\iffull evts\else events\fi({\tt lhs}), and the writes to the 
potential target objects \iffull trg\else target\fi({\tt lhs}) (determined from 
{\tt lhs} as explained below). We do not assume any order in the evaluation of 
the variables in an expression. Hence, we connect to the incoming \pos{} all 
the reads of {\tt rhs} and all the reads of {\tt lhs} except 
\iffull trg\else target\fi({\tt lhs}). We then connect each of them to the 
potential target writes to \iffull trg\else target\fi({\tt lhs}).
This is the set of objects (in the C sense~\cite{c11}) that
could be written to, according to our pointer analysis.  They are either fields
of structures, structures, arrays (independent of their offsets) or variables.
If we have \eg \verb|*(&t+y+r)=z+3| (where \texttt{t}, \texttt{y} and
\texttt{z} are shared), \texttt{t} is our target variable, and we obtain
$(R\texttt{y},W\texttt{t}) \in$ \pos{} and $(R\texttt{z},W\texttt{t}) \in$
\pos{}.
We also maintain a map from local to shared variables, to record the
data and address dependencies between abstract events. 

\par{\emph{Procedure calls} \iffull fun\else f\fi()}
We build the \pos{} corresponding to the function's body (written 
body(\iffull fun\else f\fi)) once and for all.
We then replace the call to a function \iffull fun\else f\fi() by its body. 
This ensures a better precision, in the sense that a function can be fenced 
in a given context and unfenced in another.

\par{\emph{Guarded statement}} We do not keep track of the values of the
variables at each program point. Thus we cannot evaluate the guard of a
statement.  Hence, we abstract the guard and make this statement
non-deterministically reachable, by adding a second \pos{} edge, bypassing the
statement.

\par{\emph{Forward jump to a label $L$}} We connect the previous abstract
events to the next abstract events that we generate from the program point $L$.
In \myfig\ref{tr-aeg}, we write follow($L$) for the sequence of statements
following the label $L$.

\par{\emph{Backward jump (unbounded loops)}}, \ie, a jump to a label already
visited. We %
connect the last abstract event of the copy to the first abstract
event of the original body with a \pos{} edge. 
\iffull
In \myfig\ref{tr-aeg}, begin($S$) and end($S$) are respectively the sets of the
first and last abstract events of the \pos{} sub-graph $S$. 
\fi

\iffull
\par{\emph{Assumption}} Similarly to the guarded statement, as
we cannot evaluate the condition, we abstract the assumptions by bypassing
them.
\fi

\par{\emph{Atomic sections} \!} are special goto-instructions, for modelling
idealised atomic sections without having to rely on the correctness of their
implementation. These are used in many theoretical concurrency and verification
works. For example, we use them (see \mysec\ref{sec:comparison}) for copying
data to atomic structures, as in \eg our implementation of the Chase-Lev
queue~\cite{DBLP:conf/spaa/ChaseL05}.
Reorderings cannot happen across an
atomic section, thus we place two full fences \textsf{f} right after the
beginning of the section and just before the end of section.

\par{\emph{Construction of \textsf{cmp}}}
During the construction of \pos{}, we also compute the competing pairs that
abstract external communications between threads. For each
write $w$ to a memory location $x$, we augment the $\cmp$ relation with pairs made
of this write and any write to $x$ from an interfering thread: this
abstracts the coherence \ews{}. 
Similarly, we augment $\cmp$ with pairs made of $w$ and any read of $x$ from an
interfering thread. Symmetrically, for each read $r$ of $y$, we add pairs made
of $r$ and any write to $y$ from an interfering thread to $\cmp$. This
abstracts the from-read \efr{} and read-from \rfe{} relations.
\iffull
In \myfig\ref{tr-aeg}, we use $\otimes$ to construct the $\cmp$ edges when
encountering a new interfering thread spawned, that is, when the 
goto-instruction \textsf{start\_thread} is met. We define
$\otimes$ as $A\otimes B=\{(x,y)|(x,y)\in A\times B\text{ s.t. }write(x) \lor
write(y)\}$. $\bar\emptyset$ stands for the triplet 
$(\emptyset,\emptyset,\emptyset)$.
\fi
\fi

\iffull
\else

\paragraph{Loops and arrays}

We explain how to deal with loops statically. If we build our \aeg{}
directly following the \cfg{}, with a \pos{} back-edge connecting the end of 
the body to its entry, we already handle most of the cases. %
Recall from \mysec\ref{context} that in a critical cycle (2.i) there are two
events per thread, and (2.ii) two events on the same thread target two
different locations.
Let us analyse the cases.

The first case is an iteration $i$ of this loop on which a critical cycle 
connects two events $(a_i)$ and $(b_i)$. The critical cycle will be trivially
captured by its static counterpart that abstracts in particular these events
with abstract events $(a)$ and $(b)$.

Now, for a given execution, if a critical cycle connects the event $(a_i)$ of 
an iteration $i$ to the event $(b_j)$ of a later iteration $j$ (\ie, $i\le j$),
then these events are abstracted respectively by $(a)$ and $(b)$ in the \aeg{}.
As we do not evaluate the expressions, we abstracted the loop guard and any 
local variable that would vary across the iterations. Thus, all the iterations
can be statically captured by one abstract representation of the body of the
loop. Then, thanks to the \pos{} back-edge and the transitivity of our cycle 
search, any critical cycle involving $(a_i)$ and $(b_j)$ is abstracted by a
static critical cycle relating $(a)$ and $(b)$, even though $(b)$ might be
before $(a)$ in the body of the loop.

The only case that is not handled by this approach is when $(a_i)$ and $(b_j)$
are abstracted by the same abstract event, say $(c)$. As the variables
addressed by the events on the same thread of a cycle need to be different, 
this case can only occur when $(a_i)$ and $(b_j)$ are accessing an array or a
pointer whose index or offset depends on the iteration. We do not evaluate
these offsets or indices, which implies that two accesses to two distinct array
positions might be abstracted by the same abstract event $(c)$.

In order to detect such critical cycles, we copy the body of the loop and
do not add a \pos{} back-edge. Hence, a static critical cycle will connect
$(c)$ in the first instance of the body and $(c)$ in the second instance of the
body to abstract the critical cycle involving $(a_i)$ and $(b_j)$. The 
back-edge is no longer necessary, as the abstract events reachable through this
back-edge are replicated in the second body. Thus, all the previous cases are
also covered.

We have implemented the duplication of the loop bodies only for loops that
contain accesses to arrays.  In case of nested loops, we ensure that we
duplicate each of the sub-bodies only once in order to avoid an exponential
explosion.  This approach is again sufficient owing to the maximum of two
events per thread in a critical cycle and the transitivity of \po{}.
\fi

\iffull
\paragraph{Interplay with pointer analyses}

We explain how to deal with the varying imprecision of
pointer analyses in a sound way.
We want to find all critical cycles in the static \aeg{}. Yet, some cycles
might reveal themselves only dynamically, \eg, across several iterations of
a same loop.  \myfig\ref{fig:dyn_cyc}a gives an \aeg{} for a two-thread
program, where the first thread loops and writes to a shared array
\texttt{t} and the second thread reads from this array.

This program could exhibit a \emph{message passing} pattern (see
\myfig\ref{fig:mp}), where writes in the loop could be reordered.  If we build
our \aeg{} following the \cfg{}, we obtain one abstract event for all writes of
the loop. How we address this depends on the precision of our pointer analysis.

If we have a precise pointer analysis, we insert as many abstract events as
required for the objects pointed to, as in \myfig\ref{fig:dyn_cyc}b. 
Otherwise, we underspecify the abstract event: in~\myfig\ref{fig:dyn_cyc}c,
we abstract W{\tt t[i]} by W{\tt t[*]}.  We then replicate the body of the
loop, hence the two W{\tt t[*]} in \pos{}.  They abstract any two writes to
distinct places in {\tt t} that could occur across the iterations, as
required by (2.i) and (2.ii) above.  Our tool \prog{musketeer} implements
this method.
 
If the analysis cannot determine the location of an access, we insert an
abstract event accessing any shared variable, as in \myfig\ref{fig:dyn_cyc}d. 
This event can communicate with any variable accessed in other threads.

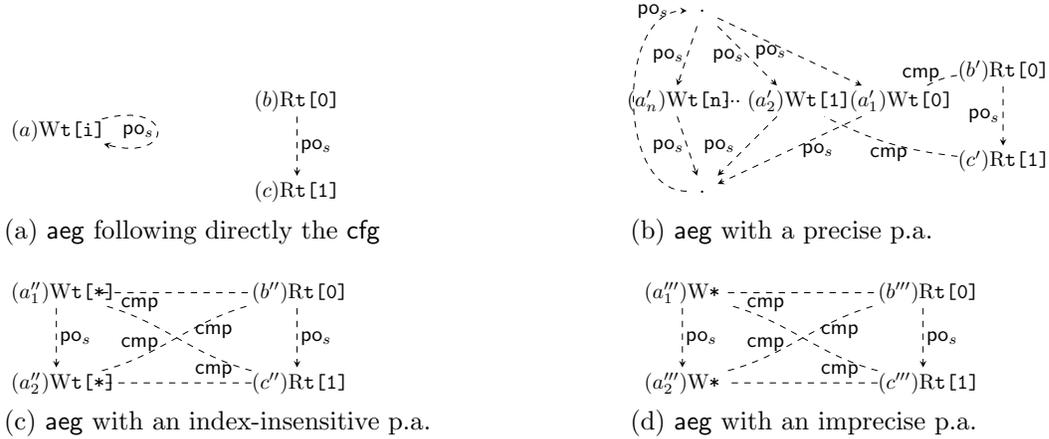
\begin{figure}[t]
\centering
\begin{tabular}{p{0.45\linewidth}p{\iffull0.003\else0.03\fi\linewidth}p{0.45\linewidth}}
  \scalebox{0.8}{
  \begin{tikzpicture}[>=stealth,thin,inner sep=0pt,text centered,shape=rectangle]

    \begin{scope}[minimum height=0.5cm,minimum width=0.5cm,text width=1.5cm]
      \node (2)  at (-2, -1.5)  {$(a)$W{\tt t[i]}};
      \node (5)  at (2, -1)  {$(b)$R{\tt t[0]}};
      \node (6)  at (2, -2.5)  {$(c)$R{\tt t[1]}};
    \end{scope}

\path[->] (2) edge [loop right, dashed] node [left=0.5mm] {\pos} (2);
\path[->] (5) edge [dashed] node [right=0.5mm] {\pos} (6);
\end{tikzpicture}}
&&
  \hspace{-0.4cm}
  \scalebox{0.8}{
  \begin{tikzpicture}[>=stealth,thin,inner sep=0pt,text centered,shape=rectangle]

    \begin{scope}[minimum height=0.5cm,minimum width=0.5cm,text width=1.5cm]
      \node (2)  at (0, -1) [text width=0.2cm] {.};
      \node (3)  at (3.2, -2.5)  {$(a_1')$W{\tt t[0]}};
      \node (4)  at (1.5, -2.5)  {$(a_2')$W{\tt t[1]}};
      \node (5)  at (0.5, -2.5)  {...};
      \node (6)  at (-0.5, -2.5)  {$(a_n')$W{\tt t[n]}};
      \node (7)  at (0, -4) [text width=0.2cm] {.};
      \node (10)  at (5, -2)  {$(b')$R{\tt t[0]}};
      \node (9)  at (5, -3.5)  {$(c')$R{\tt t[1]}};
\end{scope}

\path[->] (2) edge [dashed] node [left=0.5mm] {\pos} (3);
\path[->] (2) edge [dashed] node [left=0.5mm] {\pos} (4);
\path[->] (2) edge [dashed] node [left=0.5mm] {\pos} (6);
\path[->] (3) edge [dashed] node [right=2mm] {\pos} (7);
\path[->] (4) edge [dashed] node [left=2mm] {\pos} (7);
\path[->] (6) edge [dashed] node [left=0.5mm] {\pos} (7);
\path[->] (7) edge [dashed, bend left=90] node [above=1.5cm,right=0.5mm] {\pos} (2);
\path[->] (10) edge [dashed] node [left=0.5mm] {\pos} (9);

\path[-] (3) edge [dashed, bend left=10] node [above=0.6mm, left=0.3mm] {$\cmp$}(10);
\path[-] (9) edge [dashed, bend left=10] node [below=0.7mm] {$\cmp$}(4);

\end{tikzpicture}}
\\
(a) \aeg{} following directly the \cfg{}
&& (b) \aeg{} with a precise p.a.\\[4mm]
  \scalebox{0.8}{
  \begin{tikzpicture}[>=stealth,thin,inner sep=0pt,text centered,shape=rectangle]

    \begin{scope}[minimum height=0.5cm,minimum width=0.5cm,text width=1.5cm]
      \node (2)  at (-2, -1)  {$(a''_1)$W{\tt t[*]}};
      \node (3)  at (-2, -2.5)  {$(a''_2)$W{\tt t[*]}};
      \node (5)  at (2, -1)  {$(b'')$R{\tt t[0]}};
      \node (6)  at (2, -2.5)  {$(c'')$R{\tt t[1]}};
   \end{scope}

\path[->] (2) edge [dashed] node [right=0.5mm] {\pos} (3);
\path[->] (5) edge [dashed] node [right=0.5mm] {\pos} (6);

\path[-] (6) edge [dashed, out=165, in=-15] node [above=1mm,right=3mm] {$\cmp$}(2);
\path[-] (3) edge [dashed, out=15, in=195] node [below=1mm,left=3mm] {$\cmp$}(5);
\path[-] (6) edge [dashed] node [above=2mm,right=3mm] {$\cmp$}(3);
\path[-] (2) edge [dashed] node [below=2mm,left=3mm] {$\cmp$}(5);

\end{tikzpicture}}
&&
  \scalebox{0.8}{
  \begin{tikzpicture}[>=stealth,thin,inner sep=0pt,text centered,shape=rectangle]

    \begin{scope}[minimum height=0.5cm,minimum width=0.5cm,text width=1.5cm]
      \node (2)  at (-2, -1)  {$(a'''_1)$W{\tt *}};
      \node (3)  at (-2, -2.5)  {$(a'''_2)$W{\tt *}};
      \node (5)  at (2, -1)  {$(b''')$R{\tt t[0]}};
      \node (6)  at (2, -2.5)  {$(c''')$R{\tt t[1]}};
   \end{scope}

\path[-] (6) edge [dashed, out=165, in=-15] node [above=1mm,right=3mm] {$\cmp$}(2);
\path[-] (3) edge [dashed, out=15, in=195] node [below=1mm,left=3mm] {$\cmp$}(5);
\path[-] (6) edge [dashed] node [above=2mm,right=3mm] {$\cmp$}(3);
\path[-] (2) edge [dashed] node [below=2mm,left=3mm] {$\cmp$}(5);

\path[->] (2) edge [dashed] node [right=0.5mm] {\pos} (3);
\path[->] (5) edge [dashed] node [right=0.5mm] {\pos} (6);
\end{tikzpicture}}\\
(c) \aeg{} with an index-insensitive p.a.
&& (d) \aeg{} with an imprecise p.a.\\[2mm]
\end{tabular}
\vspace{-3mm}
\caption{Construction of \aeg{} for dynamic cycles \wrt pointer analysis 
(p.a.) precision}\label{fig:dyn_cyc}
\vspace*{-2mm}
\end{figure}

\else
\paragraph{Pointers}

We explain how to deal with the varying imprecision of pointer analyses in a
sound way.  If we have a precise pointer analysis, we insert as many
abstract events as required for the objects pointed to.  Similarly to array
accesses, a pointer might refer to two separate memory locations
dynamically, e.g., if pointer arithmetic is used.  If such an access is
detected inside a loop, the body is replicated as described above.  If the
analysis cannot determine the location of an access, we insert an abstract
event accessing any shared variable.  This event can communicate with any
variable accessed in other threads.

\fi

\paragraph{Cycle detection}
Once we have the \aeg{}, we enumerate (using
Tarjan's algorithm~\cite{DBLP:journals/siamcomp/Tarjan73}) its potential critical
cycles by searching for cycles that contain at least one edge that is a delay, as
defined in \mysec\ref{context}.

\section{Synthesis}\label{inference}

\pgfdeclarelayer{background}
\pgfsetlayers{background,main}

\begin{figure}[!t]%
\centering
\scalebox{\iffull1\else0.85\fi}{
\begin{tikzpicture}[>=stealth,thin,inner sep=0pt,text centered,shape=rectangle]

    \begin{scope}[minimum height=0.5cm,minimum width=0.5cm,text width=0.9cm]
      \node (A) at (0,0) {($c$)Rz};
      \node (E) at (0,-1.5) {($d$)Wx};
      \node (B) at (2,0) {($e$)Rx};
      \node (D) at (2,-1.5) {($f$)Ry};

      \node (C) at (4,-1.5) {($i$)Rz};
      \node (G) at (4,-3) {($j$)Wy}; %

      \node (H) at (-2,-3) {($a$)Wt}; %
      \node (I) at (-2, -4.5) {($b$)Wy};

      \node (F) at (2,-4.5) {($h$)Rt};
      \node (J) at (2, -3) {($g$)Wz};

      \node (K) at (6, -4.5) {($l$)Rz};
      \node (L) at (6, -3) {($k$)Wt};

      \node (HD) at (-2.2,-3) {};
      \node (ID) at (-2.2, -4.5) {};

      \node (FD) at (2.2,-4.5) {};
      \node (JD) at (2.2, -3) {};

      \node (KD) at (6.2, -4.5) {};
      \node (LD) at (6.2, -3) {};

      \node (DD) at (2.2,-1.5) {};

      \node (CD) at (4.2,-1.5) {};
      \node (GD) at (4.2,-3) {};

      \node (BI) at (1.8,0) {};
      \node (JI) at (1.8, -3) {};

      \node (AD) at (-0.2,0) {};
      \node (ED) at (-0.2,-1.5) {};

      \node (DI) at (1.7,-1.5) {};
      \node (FI) at (1.7,-4.5) {};
    \end{scope}

\begin{pgfonlayer}{background}
\path[->] (A) edge [dashed] node[right=0.5mm, color=white] {\pos} (E);
\path[->] (B) edge [dashed] node[left=0.5mm, color=white] {\pos} (D);
\path[->] (D) edge [dashed] node[right=0.5mm, color=white] {\pos} (J);
\path[->] (J) edge [dashed] node[left=0.5mm, color=white] {\pos} (F);
\path[->] (C) edge [dashed] node[left=0.5mm, color=white] {\pos} (G);
\path[->] (H) edge [dashed] node[right=0.5mm, color=white] {\pos} (I);
\path[<-] (K) edge [dashed] node[right=0.5mm, color=white] {\pos} (L);

\path[->] (AD) edge [%
  ] node[left=0.5mm] {\textcolor{black}{\tt dp}} 
  node[left=1cm] {cycle 1} (ED) ;
\path[->] (BI) edge [%
  ] (JI);
\path[->] (E) edge [out=30,in=210,%
  ] (B);
\path[->] (J) edge [out=30+90+45,in=210+90+45,%
  ] (A);

\path[->] (HD) edge [%
  ] node[left=0.5mm] {\textcolor{black}{\tt lwf}} 
  node[below=1.1cm,right=1cm] {cycle 3} (ID);
\path[->] (DI) edge [%
  ] (FI);
\path[->] (I) edge [out=30-30,in=210-30,%
  ] (D);
\path[->] (F) edge [out=30+90+45+15,in=210+90+45+15,%
  ] (H);

\path[<-] (D) edge [out=-45+30,in=135+30,%
  ] (G);
\path[->] (J) edge [out=45-30,in=225-30,%
  ] (C);
\path[->] (DD) edge [%
  ] (JD);
\path[->] (CD) edge [%
  ] node[right=0.5mm] {\textcolor{black}{\tt dp}} 
  node[above=1.5cm] {cycle 2} (GD);
\path[<-] (J) edge [out=-45+30,in=135+30,%
  ] (K);
\path[->] (F) edge [out=45-30,in=225-30,%
  ] (L);
\path[->] (JD) edge [%
  ] (FD);
\path[->] (LD) edge [%
  ] node[right=0.5mm] {\textcolor{black}{\tt f}}
  node[left=2.5cm,below=1.1cm] {cycle 4} (KD);
\end{pgfonlayer}
\end{tikzpicture}
}
\\[2mm]
\iffull
\fbox{
\fi
\scalebox{\iffull1\else0.75\fi}{
$
\begin{array}{l l l}
\text{\textbf{min} } 
&\multicolumn{2}{l}{\text{\tt dp}_{(e,g)} 
  + \text{\tt dp}_{(f,h)} + \text{\tt dp}_{(f,g)} + 3 \cdot (\text{\tt f}_{(e,f)}
  + \text{\tt f}_{(f,g)} + \text{\tt f}_{(g,h)}) %
  + 2 \cdot (\text{\tt lwf}_{(e,f)} 
  + \text{\tt lwf}_{(f,g)} + \text{\tt lwf}_{(g,h)})}\\
\text{\textbf{s.t.}} 
&\text{cycle 1, delay $(e,g)$:}& \text{\tt dp}_{(e,g)} + \text{\tt f}_{(e,f)} 
  + \text{\tt f}_{(f,g)}
  + \text{\tt lwf}_{(e,f)} + \text{\tt lwf}_{(f,g)} \ge 1\\
&\text{cycle 2, delay $(f,g)$:}& \text{\tt dp}_{(f,g)} + \text{\tt f}_{(f,g)} 
  + \text{\tt lwf}_{(f,g)} \ge 1\\
&\text{cycle 3, delay $(f,h)$:}& \text{\tt dp}_{(f,h)} + \text{\tt f}_{(f,g)} 
  + \text{\tt f}_{(g,h)}
  + \text{\tt lwf}_{(f,g)} + \text{\tt lwf}_{(g,h)} \ge 1\\
&\text{cycle 4, delay $(g,h)$:}& \text{\tt f}_{(g,h)} \ge 1
\end{array}
$
}
\iffull
}
\fi
\caption{Example of resolution with $\PT$}\label{fig:illustration_between}
\end{figure}%

In \myfig\ref{fig:illustration_between}, we have an \aeg{} with five threads:
$\{a,b\}, \{c,d\}$, $\{e,f,g,h\}$, $\{i,j\}$ and $\{k,l\}$.  Each node is an
abstract event computed as in the previous section. The dashed edges represent
the \pos{} between abstract events in the same thread. The full lines
represent the edges involved in a cycle. Thus the \aeg{} of
\myfig\ref{fig:illustration_between} has four potential critical cycles. We derive the
set of constraints in a process we define later in this section.
We now have a set of cycles to forbid by placing fences. Moreover, we want to
optimise the placement of the fences. 

\paragraph{Challenges}
If there is only one type of fence (as in
TSO, which only features \textsf{mfence}), optimising only consists of placing
a minimal amount of fences to forbid as many cycles as possible. For
example, placing a full fence \textsf{sync} between $f$ and $g$ in
\myfig\ref{fig:illustration_between} might forbid cycles 1, 2 and 3 under
Power, whereas placing it somewhere else might forbid at best two amongst them. 

Since we handle several types of fences for a given architecture (\eg
dependencies, \textsf{lwsync} and \textsf{sync} on Power), we can also assign
some cost to each of them. For example, following the folklore, a dependency is
less costly than an \textsf{lwsync}, which is itself less costly than a
\textsf{sync}. Given these costs, one might want to minimise their sum along
different executions: to forbid cycles 1, 2 and 3 in
\myfig\ref{fig:illustration_between}, a single \textsf{lwsync} between $f$ and
$g$ %
can be cheaper at runtime than three
dependencies respectively between $e$ and $g$, $f$ and $g$, and $f$ and $h$.
However, if we had only cycles 1 and 2, the dependencies would be cheaper.  We
see that we have to optimise both the placement and the type of fences at the
same time.

\begin{figure}%
\centering
\iffull
\fbox{
\fi
\parbox{\iffull0.76\else0.94\fi\linewidth}{%
\textbf{Input:} \aeg{} ($\mathbbm{E}_s$,\pos{},$\cmp$) and potential critical cycles \;\;$C = \{C_1, ..., C_n\}$\\
\textbf{Problem:} minimise $\sum_{(l,\textsf{t}) \in \PLACES(C)} \textsf{t}_l \times \cost(\textsf{t})$\\
\textbf{Constraints:} \text{for all $d \in \text{delays}(C)$}\\
\begin{tabular}{l}
(* for TSO, PSO, RMO, Power *)\\
\;\;if $d \in \powr$ then $\sum_{e \in \PT(d)} {\textsf f}_{e} \ge1$\\
\;\;if $d \in \poww$ then $\sum_{e \in \PT(d)} {\textsf f}_{e} + \textsf{lwf}_{e} \ge1$\\
\;\;if $d \in \porw$ then $\textsf{dp}_d + \sum_{e \in \PT(d)} {\textsf f}_{e} + \textsf{lwf}_{e}\ge1$\\
\;\;if $d \in \porr$ then $\textsf{dp}_d + \sum_{e \in \PT(d)} {\textsf f}_{e} + \textsf{lwf}_{e}\quad+ \sum_{e \in \IT(d)} \textsf{cf}_{e}\ge1$\\
(* for Power *)\\
\;\;if $d \in \cmp$ then $\sum_{e \in \CT(d)} {\textsf f}_{e}
+ \sum_{e \in \CT(d) \cap \neg\powr \cap \neg\porw}\textsf{lwf}_{e} \ge1$
\end{tabular}\\ 
\textbf{Output:} the set $\APLACES(C)$ of pairs $(l,\textsf{t})$ s.t. $\textsf{t}_l$ is set to 1 in the \ilp{} solution%
}
\iffull
}
\fi
\caption{\ilp{} for inferring fence placements}\label{mip}
\end{figure}%

We model our problem as an \emph{integer linear program} (\ilp{}) (see
\myfig\ref{mip}), which we explain in this section.
Solving
our \ilp{} gives us a set of fences to insert to forbid the cycles.  This
set of fences is optimal in that it minimises the cost function.
More precisely, the constraints are the cycles to forbid, each variable
represents a fence to insert, and the cost function sums the cost of all
fences.  %

\subsection{Cost function of the \ilp}

We handle several types of fences: full~(\textsf{f}),
lightweight~(\textsf{lwf}), control fences~(\textsf{cf}), and
dependencies~(\textsf{dp}).  On Power, the full fence is \sync, the lightweight
one \lwsync.  We write $\mathbb{T}$ for the set
$\{\textsf{dp},\textsf{f},\textsf{cf},\textsf{lwf}\}$.  We assume that each
type of fence has an \emph{a priori} cost (\eg a dependency is cheaper than a
full fence), regardless of its location in the code.  We write
$\cost(\textsf{t})$ for $\textsf{t} \in \mathbb{T}$ for this cost. 

We take as input the $\aeg$ of our program and the potential critical cycles to
fence. We
define two sets of pairs $(l,\textsf{t})$ where $l$ is a \pos{} edge of the
$\aeg{}$ and $\textsf{t}$ a type of fence.  We introduce an \ilp{} variable
$\mathsf{t}_l$ (in $\{0,1\}$) for each pair $(l,\textsf{t})$.

The set $\PLACES$ is the set of such pairs that can be inserted into the program
to forbid the cycles. The set $\APLACES$ is the set of such pairs that have
been set to $1$ by our \ilp. We output this set, as it represents the
locations in the code in need of a fence and the type of fence to insert for
each of them. We also output the total cost of all these insertions, \ie
$\sum_{(l,\textsf{t}) \in \PLACES(C)} \textsf{t}_{l}\times\cost(\textsf{t})$.
The solver should minimise this sum whilst satisfying the constraints.

\subsection{Constraints in the \ilp}

We want to forbid all the cycles in the set that we are given after filtering,
as explained in the preamble of this section. This requires placing an
appropriate fence on each delay for each cycle in this set. Different delay
pairs might need different fences, depending \eg on the directions (write or
read) of their extremities. Essentially, we follow the table in
\myfig\ref{table_fences}. For example, a write-read pair needs a full fence
(\eg \mfence{} on x86, or \sync{} on Power). A read-read pair can use anything
amongst dependencies and fences. Our constraints ensure that we use the right
type of fence for each delay pair.

\paragraph{Inequalities as constraints}
We first assume that all the program order delays are in \pos{} and we ignore
Power and ARM special features (dependencies, control fences and communication
delays). This case deals with relatively strong models, ranging from TSO to
RMO. We relax these assumptions below.

In this setting, $\PLACES(C)$ is the set of all the \pos{} delays of the cycles
in $C$. We ensure that every delay pair for every execution is fenced, by
placing a fence on the static \pos{} edge for this pair, and this for each
cycle given as input. Thus, we need at least one constraint per static delay
pair $d$ in each cycle. 

If $d$ is of the form \powr, as $(g,h)$ in \myfig\ref{fig:illustration_between}
(cycle~4), only a full fence can fix it (\cf~\myfig\ref{table_fences}), thus we
impose $\textsf{f}_d \ge 1$.  If $d$ is of the form \porr, as $(f,h)$ in
\myfig\ref{fig:illustration_between} (cycle~3), we can choose any type of
fence, \ie~$\textsf{dp}_d + \textsf{cf}_d + \textsf{lwf}_d + \textsf{f}_d \ge
1$.

Our constraints cannot be equalities because it is not certain that the
resulting system would be satisfiable. To see this, suppose our constraints
were equalities, and consider \myfig\ref{fig:illustration_between} limited to
cycles 2, 3 and 4. Using only full fences, lightweight fences, and dependencies
(\ie ignoring control fences for now), we would generate the constraints
\textbf{(i)} $\textsf{lwf}_{(f,g)}+\textsf{f}_{(f,g)}=1$ for the delay $(f,g)$
in cycle~2, \textsf{(ii)}
$\textsf{dp}_{(f,h)}+\textsf{lwf}_{(f,h)}+\textsf{f}_{(f,h)}
+\textsf{lwf}_{(g,h)}+\textsf{f}_{(g,h)}=1$ for the delay $(f,h)$ in cycle~3,
and \textbf{(iii)} $\textsf{f}_{(g,h)}=1$ for the delay $(g,h)$ in cycle~4.

Preventing the delay $(g,h)$ in cycle~4 requires a full fence, thus
$\textsf{f}_{(g,h)}=1$. By the constraint \textbf{(ii)}, and since
$\textsf{f}_{(g,h)}=1$, we derive $\textsf{f}_{(f,g)}=0$ and
$\textsf{lwf}_{(f,g)}=0$. But these two equalities are not possible given the
constraint \textbf{(i)}. By using inequalities, we allow several fences to live
on the same edge.  In fact, the constraints only ensure the soundness; the
optimality is fully determined by the cost function to minimise.

\paragraph{Delays \!\!\!} are in fact in \posplus{}, not always in
\pos{}: in \myfig\ref{fig:illustration_between}, the delay $(e,g)$ in cycle~1
does not belong to \pos{} but to \posplus{}.
Thus given a \posplus{} delay $(x,y)$, we consider all the \pos{} pairs which
appear between $x$ and $y$, \ie: $ \PT(x,y) \triangleq \{ (e_1,e_2) \in
\text{\pos{}} \mid (x,e_1) \in \text{\posstar} \land (e_2,y) \in
\text{\posstar}\}.$ For example in \myfig\ref{fig:illustration_between}, we
have $\PT(e,g)=\{(e,f),\,(f,g)\}$. %
Thus, ignoring the use of dependencies and control fences for now, for the
delay $(e,g)$ in \myfig\ref{fig:illustration_between}, we will not impose
$\textsf{f}_{(e,g)}+\textsf{lwf}_{(e,g)}\ge1$ but rather
$\textsf{f}_{(e,f)}+\textsf{lwf}_{(e,f)}+\textsf{f}_{(f,g)}
+\textsf{lwf}_{(f,g)}\ge1$.  Indeed, a full fence or a lightweight fence in
$(e,f)$ or $(f,g)$ will prevent the delay in $(e,g)$.

\paragraph{Dependencies \!\!\!\!} need more care, as they cannot
necessarily be placed anywhere between $e$ and $g$ (in the formal sense of
$\PT(e,g)$): $\textsf{dp}_{(e,f)}$ or $\textsf{dp}_{(f,g)}$ would not fix the
delay $(e,g)$, but simply maintain the pairs $(e,f)$ or $(f,g)$, leaving the
pair $(e,g)$ free to be reordered. Thus if we choose to synchronise $(e,g)$
using dependencies, we actually need a dependency from $e$ to $g$:
$\textsf{dp}_{(e,g)}$. Dependencies only apply to pairs that start with a read;
thus for each such pair (see the \porw{} and \porr{} cases in \myfig\ref{mip}),
we add a variable for the dependency: $(e,g)$ will be fixed with the constraint
$\textsf{dp}_{(e,g)}+\textsf{f}_{(e,f)}+\textsf{lwf}_{(e,f)}
+\textsf{f}_{(f,g)} +\textsf{lwf}_{(f,g)}\ge1$.

\paragraph{Control fences \!\!} placed after a conditional branch (\eg
\textsf{bne} on Power) prevent speculative reads after this branch (see
\myfig\ref{table_fences}). Thus, when building the \aeg, we built a set \poif{}
for each branch, which gathers all the pairs of abstract events such that the
first one is the last event before a branch, and the second is the first event
after that branch. We can place a control fence before the second component of
each such pair, if the second component is a read. Thus, we add $\textsf{cf}_e$
as a possible variable to the constraint for read-read pairs (see \porr{} case
in \myfig\ref{mip}, where $\IT(d) = \PT(d) \cap \poif{}$).

\paragraph{Cumulativity}
For architectures like Power, where stores are non-atomic, we need to
look for program order pairs that are connected to an external read-from
(\eg $(c,d)$ in \myfig\ref{fig:mp} has an \rf{} connected to it via
event $c$). In such cases, we need to use a \emph{cumulative fence},
\eg \lwsync{} or \sync{}, and not, for example, a dependency.

The locations to consider in such cases are: before (in \pos) the write $w$ of
the $\rfe$, or after (in \pos) the read $r$ of the $\rfe$, \ie $\CT(w,r) = \{
(e_1,e_2) \mid (e_1,e_2) \in \text{\pos} \land ((e_2,w) \in \text{\posstar} \lor (r,e_1)
\in \text{\posstar}) \}$. In \myfig\ref{fig:illustration_between} (cycle
2), $(g,i)$ over-approximates an \rfe{} edge, and the edges where we can insert
fences are in $\CT(g,i)=\{(f,g),\,(i,j)\}$.

We need a cumulative fence as soon as there is a potential $\rfe$,
even if the adjacent \pos{} pairs do not form a delay. For example in
\myfig\ref{fig:mp}, suppose there is a dependency between the reads on $T_1$,
and a fence maintaining write-write pairs on $T_0$. In that case we need to
place a cumulative fence to fix the $\rfe$, even if the two \pos{} pairs are
themselves fixed. 
Thus, we quantify over all \pos{} pairs when we need to place cumulative
fences.  As only \textsf{f} and \textsf{lwf} are cumulative, we have 
$\PLACES(C)\triangleq$ {\small 
$\left\{ (l,t) \mid (t\in\{\textsf{dp}\}\right.$ 
$\land~l\in\text{delays}(C))$ 
$\lor (t\in \mathbb{T}\backslash\{\textsf{dp}\}$ 
$\land~l\in \bigcup_{d\in\text{delays}(C)}\PT(d))$ 
$\lor (t\in\{\textsf{f},\textsf{lwf}\}$
$\left.\land~l\in\text{\pos}(C))\right\}$.}  
\iffull
\subsection{Comparison with \textsf{trencher}}
\else
\paragraph{Comparison with \textsf{trencher}}
\fi
\begin{wrapfigure}{r}{0.45\linewidth}
  \centering
  \iffull
  \vspace{-0.4cm}
  \else
  \vspace{-0.8cm}
  \fi
  \scalebox{0.9}{
  \begin{tikzpicture}[>=stealth,thin,inner sep=0pt,text centered,shape=rectangle]

    \begin{scope}[minimum height=0.5cm,minimum width=0.5cm,text width=1.5cm]
      \node (2)  at (-3, -0.5)  {$(a)$W{\tt x}};
      \node (3)  at (-3, -2.5)  {$(b)$R{\tt y}};
      \node (5)  at (0, 0)  {$(c)$W{\tt y}};
      \node (7)  at (0, -1.5)  {$(e)$};
      \node (8)  at (-0.5, -1.5)  {$(d)$};
      \node (9)  at (0.5, -1.5)  {$(f)$};
      \node (6)  at (0, -3)  {$(g)$R{\tt x}};
   \end{scope}

\path[->] (2) edge [dashed, bend right=10] node [right=0.5mm] {\textsf{f}} (3);
\path[->] (5) edge [dashed] (7);
\path[->] (7) edge [dashed] (6);
\path[->] (5) edge [dashed, bend right=10] (8);
\path[->] (8) edge [dashed, bend right=10] (6);
\path[->] (5) edge [dashed, bend left=10] node [right=0.5mm] {\pos} (9);
\path[->] (9) edge [dashed, bend left=10] node [right=0.5mm] {\pos} (6);
\path[-] (6) edge [dashed, out=165, in=-15] node [above=1cm] {$\cmp$} (2);
\path[-] (3) edge [dashed, out=15, in=195] node [below=1cm] {$\cmp$} (5);

\end{tikzpicture}}
\vspace*{-2mm}
\caption{Cycles sharing the edge $(a,b)$}\label{fig:2sb}
\vspace*{-5mm}
\end{wrapfigure}
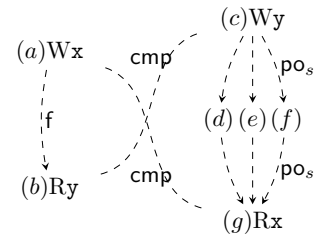

We illustrate the difference between
\prog{trencher}~\cite{DBLP:conf/esop/BouajjaniDM13} and our approach using
\myfig\ref{fig:2sb}.  There are three cycles that share the edge $(a,b)$. They
differ in the path taken between nodes $c$ and $g$. Suppose that the user has
inserted a full fence between $a$ and $b$. To forbid the three cycles, we need
to fence the thread on the right.

The \prog{trencher} algorithm first calculates which pairs can be reordered:
in our example, these are $(c,g)$ via $d$, $(c,g)$ via $e$ and $(c,g)$ via
$f$.  It then determines at which locations a fence could be placed.  In our
example, there are $6$ options: $(c,d)$, $(d,g)$, $(c,e)$, $(e,g)$, $(c,f)$,
and $(f,g)$.  The encoding thus uses $6$ variables for the fence locations. 
The algorithm then gathers all the \emph{irreducible} sets of locations to
be fenced to forbid the delay between $c$ and $g$, where ``irreducible''
means that removing any of the fences would prevent this set from fully
fixing the delay.  As all the paths that connect $c$ and $g$ have to be
covered, \prog{trencher} needs to collect all the combinations of one fence
per path.  There are $2$ locations per path, leading to $2^3$ sets. 
Consequently, as stated in~\cite{DBLP:conf/esop/BouajjaniDM13},
\prog{trencher} needs to construct an exponential number of sets.

Each set is encoded in the \ilp{} with one variable. For this example,
\prog{trencher} thus uses $6+8$ variables. It also generates one constraint
per delay (here, $1$) to force the solver to pick a set, and $8$ constraints
to enforce that all the location variables are set to $1$ if the set
containing these locations is picked.

By contrast, \prog{musketeer} only needs $6$ variables: the possible
locations for fences.  We detect three cycles, and generate only three
constraints to fix the delay.  Thus, on a parametric version of the example,
\prog{trencher}'s \ilp{} grows exponentially whereas \prog{musketeer}'s is
linear-sized.

\iffull
\section{Implementation, Experiments and Impact}
\else
\section{Implementation and Experiments}
\fi
\label{sec:comparison}

\begin{figure}[!t]
\begin{center}
\hspace*{-2.2mm}
\scalebox{\iffull0.8\else0.94\fi}{
\begin{tabular}{l|cccccccccc|cccccccccccc}
\multicolumn{1}{l}{}& \multicolumn{10}{c}{\textsc{classic}} & \multicolumn{12}{c}{\textsc{fast}}\\
& \multicolumn{2}{c}{Dek} &  \multicolumn{2}{c}{Pet} &  \multicolumn{2}{c}{Lam} &  \multicolumn{2}{c}{Szy} &  \multicolumn{2}{c}{Par} & \multicolumn{2}{c}{Cil} & \multicolumn{2}{c}{CL} & \multicolumn{2}{c}{Fif} & \multicolumn{2}{c}{Lif} & \multicolumn{2}{c}{Anc} & \multicolumn{2}{c}{Har}\\
\hline
LoC & \multicolumn{2}{c}{50} & \multicolumn{2}{c}{37} & \multicolumn{2}{c}{72} & \multicolumn{2}{c}{54} & \multicolumn{2}{c}{96} & \multicolumn{2}{c}{97} & \multicolumn{2}{c}{111} & \multicolumn{2}{c}{150} & \multicolumn{2}{c}{152} & \multicolumn{2}{c}{188} & \multicolumn{2}{c}{179}\\
\hline
\prog{dfence} & -- & -- & -- & -- & -- & -- & -- & -- & -- & -- & 7.8 & 3 & 6.2 & 3 & $\sim$ & 0 & $\sim$ & 0 & $\sim$ & 0 & $\sim$ & 0\\
\prog{memorax} & 0.4 & 2 & 1.4 & 2 & 79.1 & 4 & -- & -- & -- & -- & -- & -- & -- & -- & -- & -- & -- & -- & -- & -- & -- & --\\
\textbf{\prog{musketeer}} & \textbf{0.0} & \textbf{5} & \textbf{0.0} & \textbf{3} & \textbf{0.0} & \textbf{8} & \textbf{0.0} & \textbf{8} & \textbf{0.0} & \textbf{3} & \textbf{0.0} & \textbf{3} & \textbf{0.0} & \textbf{1} & \textbf{0.1} & \textbf{1} & \textbf{0.0} & \textbf{1} & \textbf{0.1} & \textbf{1} & \textbf{0.6} & \textbf{4}\\
\prog{offence} & 0.0 & 2 & 0.0 & 2 & 0.0 & 8 & 0.0 & 8 & -- & -- & -- & -- & -- & -- & -- & -- & -- & -- & -- & -- & -- & --\\
\prog{pensieve} & 0.0 & 16 & 0.0 & 6 & 0.0 & 24 & 0.0 & 22 & 0.0 & 7 & 0.0 & 14 & 0.0 & 8 & 0.1 & 33 & 0.0 & 29 & 0.0 & 44 & 0.1 & 72\\
\prog{remmex} & 0.5 & 2 & 0.5 & 2 & 2.0 & 4 & 1.8 & 5 & -- & -- & -- & -- & -- & -- & -- & -- & -- & -- & -- & -- & -- & --\\
\prog{trencher} & 1.6 & 2 & 1.3 & 2 & 1.7 & 4 & -- & -- & 0.5 & 1 & 8.6 & 3 & -- & -- & -- & -- & -- & --& -- & --& -- & --\\
\end{tabular}
}\end{center}
\vspace{-5mm}
\caption{All tools on the \textsc{classic} and \textsc{fast} series for TSO}\label{fast}\label{class}
\end{figure}

We implemented our new method, in addition to all the methods described in
\mysec\ref{sec:motivation}, in our tool \prog{musketeer}, using \prog{glpk}
(\url{http://www.gnu.org/software/glpk}) as the \ilp{} solver.
\iffull
\prog{musketeer} is a completely automated source-to-source transformation
for concurrent C program. Once the locations and types of fences have been 
inferred, the insertion in the source itself is performed by a script. This 
step is relatively straightforward for memory fences. Inserting dependencies 
in C is more challenging, due to the multiple optimisations that the compiler 
will perform. We explain how we address this issue in 
\mysec\ref{sec:insertion_practice}.\\\indent
\fi
We compare 
\iffull our method and the methods we reimplemented \else these methods\fi to 
the existing tools listed in \mysec\ref{related-work}
\iffull
on classic examples from literature and some Debian executables in 
\mysec\ref{sec:experiments}. We finally check the impact on runtime of the 
fences inferred and inserted in \prog{memcached}, a Debian executable of about
10000 lines of code.
\else
. 
\fi

\iffull
\subsection{Inserting synchronisation in practice}
\label{sec:insertion_practice}
\fi

\iffull

\begin{figure}[b] \centering 
\scalebox{1}{
     \begin{tabular}{l} r1 = x;\\ r2 = r1+2;\\ r3 = y;\\
     \phantom{\_}
     \end{tabular}
} \hspace{0.9cm} 
\scalebox{1}{
     \begin{tabular}{l} r1 = x;\\ \_\_asm\_\_ ("mfence");\\ r2 = r1+2;\\ r3 =
y;\\ \phantom{\_\_asm\_\_ ("mfence");} \end{tabular}
    } 
\hspace{0.9cm} 
\scalebox{1}{
     \begin{tabular}{l} r1 = x;\\ r2 = r1+2;\\ \_\_asm\_\_ ("mfence");\\ r3 =
y;\\ \phantom{\_\_asm\_\_ ("mfence");} \end{tabular}
    } %
\vspace{-2mm}
\caption{Choices for placing a fence.}
\label{fig:code}
\end{figure}

\begin{figure}[]
\centering
\begin{tabular}{m{.35\linewidth} m{.2\linewidth}}
\begin{lstlisting}[showlines=true, numberblanklines=true]
void* t0(void* arg) {
   int r1;
   while(x);
   asm volatile ("isb");
   r1 = y;
}





\end{lstlisting}
&
\begin{lstlisting}
t0:
   ldr r2,.L5
.L2:
   ldr r3,[r2,#0]
   cmp r3,#0
   bne .L2
   isb
   ldr r3,.L5+4
   ldr r3,[r3,#0]
   bx  lr
.L5:
\end{lstlisting}
\end{tabular}
\vspace*{-4mm}
\caption{\small{\texttt{isb.c}} and \small{\texttt{isb-O3.s}}.}
\label{fig:while}
\end{figure}

\noindent
Given an \aeg, we return the static program order edges~where we should place a
fence to forbid the critical cycles. Then we have some freedom for the
fence placement in the actual code.
Consider \eg the program on the left of~\myfig\ref{fig:code}. The corresponding
\aeg{} is $(R x,R y) \in \text{\pos{}}$. To fence this edge, we can place a
fence either as in the middle or as on the right of \myfig\ref{fig:code}, namely just
after the first component of a delay pair, or just before the last. Our tool
offers these two options.
We next illustrate how we concretely insert fences and dependencies in a piece of C code.
\paragraph{Fences}
are all handled the same way; we simply inline
an assembly fence. 
For example, for a read-read pair separated by a branch (lines 3 and 5 in
\myfig\ref{fig:while} on the left), we can insert a control fence, \eg \isb{} on ARM. The
compiler keeps the fence in place, as one can see in the compiled code in \myfig\ref{fig:while} on
the right.
The while loop (including the read of $x$) is implemented by lines 3 to 6, then comes the \isb{} (line 7),
and the read of $y$ corresponds to lines 8 and 9.

\paragraph{Dependencies}
require us to rewrite the code.  Consider
a read-read pair, corresponding to lines 3 and 9 on the left of \myfig\ref{fig:dp}.
We enforce an \emph{address dependency} from the read of x to the read of y, by
using a register (r3) to perform some computation which always returns $0$ (in
this case XOR-ing a register with itself),
then add this result to the address of y.
Again, the compiler does not optimise
this dependency (see lines 4 to 8 on the right of \myfig\ref{fig:dp}).

\begin{figure}[b]
\vspace{-0.35cm}\centering
\begin{tabular}{m{.4\linewidth} m{.3\linewidth}}
\begin{lstlisting}[showlines=true, numberblanklines=true]
void* t0(void* arg) {
   int r1, r2;
   r1=x;
   int r3;
   asm volatile (
     "eors %0, %1, %1"
     :"=r"(r3) 
     :"r"(r1));
   r2=*(&y+r3);
}
\end{lstlisting}
&
\begin{lstlisting}[showlines=true, numberblanklines=true]
t0:
  movw r3,#:lower16:x
  movt r3,#:upper16:x
  ldr  r2,[r3, #0]
  eors r2, r2,r2
  movw r3,#:lower16:y
  movt r3,#:upper16:y
  ldr  r3,[r3,r2,lsl #2]
  bx   lr

\end{lstlisting}
\end{tabular}
\vspace*{-6mm}
\caption{\small{\texttt{addr.c}} and \small{\texttt{addr-O3.s}}.}
\label{fig:dp}
\vspace*{-3mm}
\end{figure}

\fi

\iffull
\subsection{Experiments and benchmarks}
\label{sec:experiments}
\fi

Our tool analyses C programs.  \prog{dfence} also handles C code, but
requires some high-level specification for each program, which was not
available to us.  \prog{memorax} works on a process-based language that is
specific to the tool.  \prog{offence} works on a subset of assembler for
x86, ARM and Power.  \prog{pensieve} originally handled Java, but we did not
have access to it and have therefore re-implemented the method. 
\prog{remmex} handles Promela-like programs.  \prog{trencher} analyses
transition systems.  Most of the tools come with some of the benchmarks in
their own languages; not all benchmarks were however available for each
tool.  We have re-implemented some of the benchmarks for \prog{offence}.

We now detail our experiments. 
\textsc{classic} and \textsc{fast} gather examples from the
literature and related work.
The \textsc{debian} benchmarks are packages of Debian Linux 7.1.
\textsc{classic} and \textsc{fast} were run on a x86-64 Intel
Core2 Quad Q9550 machine with 4 cores (2.83\,GHz) and 4\,GB of RAM.  \textsc{debian}
was run on a x86-64 Intel 
Core i5-3570 machine with 4 cores (3.40\,GHz) and
4\,GB of RAM.
\paragraph{\textsc{classic} \!\!\!} consists of Dekker's mutex
(Dek)~\cite{DBLP:journals/cacm/Dijkstra65}; Peterson's mutex
(Pet)~\cite{DBLP:journals/ipl/Peterson81}; Lamport's fast mutex
(Lam)~\cite{DBLP:journals/tocs/Lamport87}; Szymanski's mutex
(Szy)~\cite{DBLP:conf/ics/Szymanski88}; and Parker's bug (Par)~\cite{Dice09}.
We ran all tools in this series for TSO (the model common to all).
For each
example, \myfig\ref{class} gives the number of fences inserted, and the time
(in sec) needed. When an example is not available in the input language of
a tool,
we write ``--''.
The first four tools place fences to enforce stability/robustness~\cite{am11,bmm11}; the
last three to satisfy a given safety property.
We used \prog{memorax} with the option {\tt -o1}, to compute one
\emph{maximal permissive} set and not all.  For \prog{remmex} on Szymanski,
we give the number of fences found by default (which may be non-optimal). 
Its ``maximal permissive'' option lowers the number to $2$, at the cost of a
slow enumeration.
As expected, \prog{musketeer} is less precise than most tools, but
outperforms all of them.

\paragraph{\textsc{fast} \!\!\!} gathers Cil, Cilk 5 Work Stealing Queue
(WSQ)~\cite{DBLP:conf/pldi/FrigoLR98}; CL, Chase-Lev
WSQ~\cite{DBLP:conf/spaa/ChaseL05}; Fif, Michael et \al{}'s FIFO
WSQ~\cite{DBLP:conf/ppopp/MichaelVS09}; Lif, Michael et \al{}'s LIFO
WSQ~\cite{DBLP:conf/ppopp/MichaelVS09}; Anc, Michael et \al{}'s Anchor
WSQ~\cite{DBLP:conf/ppopp/MichaelVS09}; Har, Harris'
set~\cite{DBLP:conf/wdag/DetlefsFGMSS00}.
For each example and tool, \myfig\ref{fast} gives the number of fences inserted
(under TSO) and the time needed to do so.
For \prog{dfence}, we used the setting of~\cite{DBLP:conf/pldi/LiuNPVY12}:
the tool has up to $20$ attempts to find fences.  We were unable to apply
\prog{dfence} on some of the \textsc{fast} examples: we thus reproduce the
number of fences given in~\cite{DBLP:conf/pldi/LiuNPVY12}, and write $\sim$
for the time.
We applied \prog{musketeer} to this series, for all architectures. The fencing
times for TSO and Power are almost identical, except for the largest
example, namely Har ($0.1\,s$ vs $0.6\,s$).

\begin{figure}[!t]
\begin{center}
\scalebox{0.9}{
\begin{tabular}{l|@{\qquad}r@{\qquad}r@{\qquad}|@{\qquad}r@{\qquad}r@{\qquad}|@{\qquad}r@{\qquad}r@{\qquad}}
& \multicolumn{2}{c}{} &\multicolumn{2}{c@{\qquad\qquad}}{TSO} & \multicolumn{2}{c@{\qquad\qquad}}{Power} \\
 & LoC& nodes & fences & time & fences & time\\
\hline
\prog{memcached} & 9944 & 694 & 3 & 13.9s  & 70 & 89.9s \\ %
\prog{lingot} & 2894 & 183 & 0 & 5.3s & 5 & 5.3s \\ %
\prog{weborf} & 2097 & 73 & 0 & 0.7s & 0 & 0.7s\\ %
\prog{timemachine} & 1336 & 129  & 2 & 0.8s & 16 & 0.8s\\ %
\prog{see} & 2626 & 171 & 0 & 1.4s & 0 & 1.5s \\ %
\prog{blktrace} & 1567 & 615 & 0 & 6.5s & -- & timeout\\
\prog{ptunnel} & 1249 & 1867 & 2 & 95.0s & -- & timeout \\ %
\prog{proxsmtpd} & 2024 & 10 & 0 & 0.1s & 0 & 0.1s \\ %
\prog{ghostess} & 2684 & 1106 & 0 & 25.9s & 0 & 25.9s \\ %
\prog{dnshistory} & 1516 & 1466 & 1 & 29.4s & 9 & 64.9s \\ %
\end{tabular}
}
\end{center}
\caption{\prog{musketeer} on selected benchmarks in \textsc{debian} series for TSO and Power}\label{debian}
\end{figure}

\iffull
\begin{figure}[b]
\centering

\subfigure{
\includegraphics[scale=1]{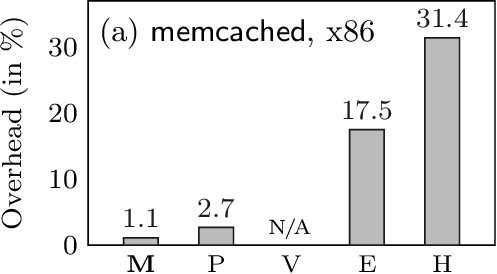}
}
\subfigure{
\includegraphics[scale=1]{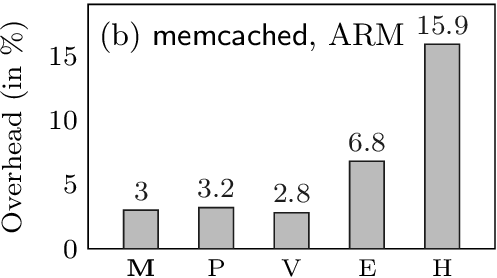}
}
\caption{Runtime overheads due to inserted fences in \prog{memcached} for
each strategy}
\label{fig:performance-x86app}
\iffull
\else
\vspace*{-4mm}
\fi
\end{figure}
\fi

\paragraph{\textsc{debian} \!\!\!} gathers $374$ executables.  These are a
subset of the goto-programs that have been built from packages of Debian
Linux~7.1 by Michael Tautschnig.  A~small excerpt of our results is given
in \myfig\ref{debian}.  The full data set, including a comparison with the
methods from Sec.~\ref{sec:motivation}, is provided at
\url{http://www.cprover.org/wmm/musketeer}.  For each program, we give the
lines of code and number of nodes in the \aeg{}.
We used \prog{musketeer} on these programs to demonstrate its scalability
and its ability to handle deployed code.  Most programs already contain
fences or operations that imply them, such as compare-and-swaps or locks. 
Our tool \prog{musketeer} takes these fences into account and infers a set
of additional fences sufficient to guarantee SC.
The largest program we handle is \prog{memcached} ($\sim10000$ LoC).  Our tool needs
$13.9\,s$ to place fences for TSO, and $89.9\,s$ for Power. 
A more meaningful
measure for the hardness of an instance is the number of nodes in the \aeg{}.
For example, \prog{ptunnel} has 1867 nodes and 1249 LoC.  The fencing takes
$95.0\,s$ for TSO, but times out for Power due to the number of cycles.
\iffull
Not all fences inferred by
\prog{musketeer} are necessary to enforce SC, due to the imprecision introduced
by the \aeg{} abstraction. However, as Section~\ref{sec:motivation} shows, the
execution time overhead of the program versions with fences inserted by \prog{musketeer}
is still very low.

\fi

\iffull
\subsection{Impact on runtime}

We finally measure the impact of fences for the program \prog{memcached}, 
running experiments that are similar to those in \mysec\ref{sec:motivation}.
We built new versions of \prog{memcached} according to the fencing 
strategies described in \mysec\ref{sec:motivation}. We used in particular 
the \prog{memtier} benchmarking tool to generate a workload for the 
\prog{memcached} daemon, and killed the daemon after $60\,s$. 
\myfig\ref{fig:performance-x86} shows the mean overhead \wrt{} the original,
unmodified program. 

Unsurprisingly, adding a fence after \emph{every} access to static or heap data
has a significant performance effect. Similarly, adding fences via an escape
analysis is expensive, yielding overheads of up to 17.5\,\%. Amongst the
approaches guaranteeing SC (\ie all but~\textsc{v}), the best results were
achieved with our tool \prog{musketeer}. We again computed the confidence 
intervals to check the statistical significance in \myfig\ref{fig:ds-exp}.

\begin{figure}[]
\centering
\begin{tabular}{l|c|c|c|c}
& \prog{stack} on x86 & \prog{stack} on ARM & \prog{queue} on x86 & \prog{queue} on ARM\\
\hline
\textsc{(o)} & [9.757; 9.798] & [11.291; 11.369] & [11.947; 11.978] & [20.441; 20.634]\\
\textsc{(m)} & [9.818; 9.850] & [11.316; 11.408] & [12.067; 12.099] & [20.687; 20.857]\\
\textsc{(p)} & [10.077; 10.155] & [11.995; 12.109] & [13.339; 13.373] & [22.035; 22.240]\\
\textsc{(v)} &        N/A       & [11.779; 11.834] &       N/A        & [21.334; 21.526]\\
\textsc{(e)} & [11.316; 11.360] & [13.071; 13.200] & [13.949, 13.981] & [22.722; 22.903]\\
\textsc{(h)} & [12.286; 12.325] & [14.676; 14.844] & [14.941, 14.963] & [25.468; 25.633]
\end{tabular}
\caption{Confidence intervals for data structure experiments}
\label{fig:ds-exp}
\end{figure}
\fi

\vspace{-2mm}
\section{Conclusion}\label{conclusion}

We introduced a novel method for deriving a set of fences, which we
implemented in a new tool called \prog{musketeer}.  We compared it to
existing tools and observed that it outperforms them.  We demonstrated on
our \textsc{debian} series that \prog{musketeer} can handle deployed code,
with a large potential for scalability.

\paragraph{Acknowledgements} 

We thank Michael Tautschnig for the Debian binaries,  Mohamed Faouzi Atig, Egor
Derevenetc, Carsten Fuhs, Alexander Linden, Roland Meyer, Tyler Sorensen,
Martin Vechev, Eran Yahav and our reviewers for their feedback. We thank
Alexander Linden and Martin Vechev again for giving us access to their tools.

{
\iffull
\bibliographystyle{amsalpha}
\else
\bibliographystyle{splncs03}
\fi
\bibliography{bibliography,jade}

\newcommand{\etalchar}[1]{$^{#1}$}
\providecommand{\bysame}{\leavevmode\hbox to3em{\hrulefill}\thinspace}
\providecommand{\MR}{\relax\ifhmode\unskip\space\fi MR }
\providecommand{\MRhref}[2]{%
  \href{http://www.ams.org/mathscinet-getitem?mr=#1}{#2}
}
\providecommand{\href}[2]{#2}
\begin{thebibliography}{MSM{\etalchar{+}}11}

\bibitem[AAC{\etalchar{+}}13]{DBLP:conf/tacas/AbdullaACLR13}
Parosh~Aziz Abdulla, Mohamed~Faouzi Atig, Yu-Fang Chen, Carl Leonardsson, and
  Ahmed Rezine, \emph{Memorax, a precise and sound tool for automatic fence
  insertion under {TSO}}, Tools and Algorithms for the Construction and
  Analysis of Systems (TACAS), LNCS, Springer, 2013, pp.~530--536.

\bibitem[AG95]{ag96}
S.~V. Adve and K.~Gharachorloo, \emph{Shared memory consistency models: {A}
  tutorial}, IEEE Computer \textbf{29} (1995), no.~12, 66--76.

\bibitem[AKL{\etalchar{+}}11]{alg11}
Jade Alglave, Daniel Kroening, John Lugton, Vincent Nimal, and Michael
  Tautschnig, \emph{Soundness of data flow analyses for weak memory models},
  Programming Languages and Systems (APLAS), LNCS, vol. 7078, Springer, 2011,
  pp.~272--288.

\bibitem[AKNT13]{DBLP:conf/esop/AlglaveKNT13}
Jade Alglave, Daniel Kroening, Vincent Nimal, and Michael Tautschnig,
  \emph{Software verification for weak memory via program transformation},
  European Symposium on Programming (ESOP), 2013.

\bibitem[AM11]{am11}
J.~Alglave and L.~Maranget, \emph{Stability in weak memory models}, Computer
  Aided Verification (CAV), LNCS, vol. 6806, Springer, 2011, pp.~50--66.

\bibitem[AMSS10]{ams10}
J.~Alglave, L.~Maranget, S.~Sarkar, and P.~Sewell, \emph{Fences in weak memory
  models}, Computer Aided Verification (CAV), LNCS, vol. 6174, Springer, 2010,
  pp.~258--272.

\bibitem[BDM13]{DBLP:conf/esop/BouajjaniDM13}
Ahmed Bouajjani, Egor Derevenetc, and Roland Meyer, \emph{Checking and
  enforcing robustness against {TSO}}, European Symposium on Programming
  (ESOP), LNCS, vol. 7792, Springer, 2013, pp.~533--553.

\bibitem[BMM11]{bmm11}
A.~Bouajjani, R.~Meyer, and E.~Moehlmann, \emph{Deciding robustness against
  total store ordering}, Automata, Languages and Programming (ICALP), LNCS,
  vol. 6756, Springer, 2011, pp.~428--440.

\bibitem[c1111]{c11}
\emph{Information technology -- {P}rogramming languages -- {C}}, BS ISO/IEC
  9899:2011, 2011.

\bibitem[CL05]{DBLP:conf/spaa/ChaseL05}
David Chase and Yossi Lev, \emph{Dynamic circular work-stealing deque}, SPAA,
  ACM, 2005, pp.~21--28.

\bibitem[DFG{\etalchar{+}}00]{DBLP:conf/wdag/DetlefsFGMSS00}
David Detlefs, Christine~H. Flood, Alex Garthwaite, Paul~A. Martin, Nir Shavit,
  and Guy~L. Steele~Jr., \emph{Even better {DCAS}-based concurrent deques},
  Distributed Computing (DISC), LNCS, vol. 1914, Springer, 2000, pp.~59--73.

\bibitem[Dic09]{Dice09}
David Dice, November 2009.

\bibitem[Dij65]{DBLP:journals/cacm/Dijkstra65}
Edsger~W. Dijkstra, \emph{Solution of a problem in concurrent programming
  control}, Commun. ACM \textbf{8} (1965), no.~9, 569.

\bibitem[FLM03]{DBLP:conf/ics/FangLM03}
Xing Fang, Jaejin Lee, and Samuel~P. Midkiff, \emph{Automatic fence insertion
  for shared memory multiprocessing}, International Conference on
  Supercomputing (ICS), ACM, 2003, pp.~285--294.

\bibitem[FLR98]{DBLP:conf/pldi/FrigoLR98}
Matteo Frigo, Charles~E. Leiserson, and Keith~H. Randall, \emph{The
  implementation of the {C}ilk-5 multithreaded language}, PLDI, ACM, 1998,
  pp.~212--223.

\bibitem[KSKZ09]{DBLP:conf/sigsoft/KahlonSKZ09}
Vineet Kahlon, Nishant Sinha, Erik Kruus, and Yun Zhang, \emph{Static data race
  detection for concurrent programs with asynchronous calls}, FSE, 2009.

\bibitem[KVY10]{DBLP:conf/fmcad/KupersteinVY10}
Michael Kuperstein, Martin~T. Vechev, and Eran Yahav, \emph{Automatic inference
  of memory fences}, Formal Methods in Computer-Aided Design (FMCAD), IEEE,
  2010, pp.~111--119.

\bibitem[KVY11]{DBLP:conf/pldi/KupersteinVY11}
\bysame, \emph{Partial-coherence abstractions for relaxed memory models}, PLDI,
  2011, pp.~187--198.

\bibitem[KY96]{DBLP:journals/jpdc/KrishnamurthyY96}
Arvind Krishnamurthy and Katherine~A. Yelick, \emph{Analyses and optimizations
  for shared address space programs}, J. Par. Dist. Comp. \textbf{38} (1996),
  no.~2.

\bibitem[Lam79]{lam79}
L.~Lamport, \emph{How to {M}ake a {C}orrect {M}ultiprocess {P}rogram {E}xecute
  {C}orrectly on a {M}ultiprocessor}, IEEE Trans. Comput. \textbf{46} (1979),
  no.~7.

\bibitem[Lam87]{DBLP:journals/tocs/Lamport87}
Leslie Lamport, \emph{A fast mutual exclusion algorithm}, ACM Trans. Comput.
  Syst. \textbf{5} (1987), no.~1.

\bibitem[LNP{\etalchar{+}}12]{DBLP:conf/pldi/LiuNPVY12}
Feng Liu, Nayden Nedev, Nedyalko Prisadnikov, Martin~T. Vechev, and Eran Yahav,
  \emph{Dynamic synthesis for relaxed memory models}, PLDI, ACM, 2012,
  pp.~429--440.

\bibitem[LP01]{lp01}
J.~Lee and D.A. Padua, \emph{Hiding relaxed memory consistency with a
  compiler}, IEEE Transactions on Computers \textbf{50} (2001), 824--833.

\bibitem[LW13]{DBLP:conf/tacas/LindenW13}
Alexander Linden and Pierre Wolper, \emph{A verification-based approach to
  memory fence insertion in {PSO} memory systems}, Tools and Algorithms for the
  Construction and Analysis of Systems (TACAS), LNCS, vol. 7795, Springer,
  2013, pp.~339--353.

\bibitem[MSM{\etalchar{+}}11]{DBLP:conf/pldi/MarinoSMMN11}
Daniel Marino, Abhayendra Singh, Todd~D. Millstein, Madanlal Musuvathi, and
  Satish Narayanasamy, \emph{A case for an {SC}-preserving compiler}, PLDI,
  ACM, 2011, pp.~199--210.

\bibitem[MVS09]{DBLP:conf/ppopp/MichaelVS09}
Maged~M. Michael, Martin~T. Vechev, and Vijay~A. Saraswat, \emph{Idempotent
  work stealing}, Principles and Practice of Parallel Programming (PPOPP), ACM,
  2009, pp.~45--54.

\bibitem[ND13]{DBLP:conf/oopsla/NorrisD13}
Brian Norris and Brian Demsky, \emph{{CDS}checker: checking concurrent data
  structures written with {C/C++} atomics}, Object Oriented Programming Systems
  Languages \& Applications (OOPSLA), 2013, pp.~131--150.

\bibitem[OSS09]{oss09}
S.~Owens, S.~Sarkar, and P.~Sewell, \emph{A better x86 memory model:
  x86-{TSO}}, Theorem Proving in Higher Order Logics (TPHOLs), LNCS, vol. 5674,
  Springer, 2009, pp.~391--407.

\bibitem[Pet81]{DBLP:journals/ipl/Peterson81}
Gary~L. Peterson, \emph{Myths about the mutual exclusion problem}, Inf.
  Process. Lett. \textbf{12} (1981), no.~3, 115--116.

\bibitem[ppc09]{ppc:2.06}
\emph{Power isa version 2.06}, 2009.

\bibitem[SFW{\etalchar{+}}05]{DBLP:conf/ppopp/SuraFWMLP05}
Zehra Sura, Xing Fang, Chi-Leung Wong, Samuel~P. Midkiff, Jaejin Lee, and
  David~A. Padua, \emph{Compiler techniques for high performance sequentially
  consistent {J}ava programs}, PPOPP, ACM, 2005, pp.~2--13.

\bibitem[spa94]{sparc:94}
\emph{{S}parc {A}rchitecture {M}anual {V}ersion 9}, 1994.

\bibitem[SS88]{ss88}
D.~Shasha and M.~Snir, \emph{Efficient and correct execution of parallel
  programs that share memory}, TOPLAS \textbf{10} (1988), no.~2, 282--312.

\bibitem[Szy88]{DBLP:conf/ics/Szymanski88}
Boleslaw~K. Szymanski, \emph{A simple solution to {L}amport's concurrent
  programming problem with linear wait}, ICS, 1988, pp.~621--626.

\bibitem[Tar73]{DBLP:journals/siamcomp/Tarjan73}
R.~Tarjan, \emph{Enumeration of the elementary circuits of a directed graph},
  SIAM J. Comput. \textbf{2} (1973), no.~3, 211--216.

\bibitem[VZN11]{DBLP:conf/sas/VafeiadisN11}
Viktor Vafeiadis and Francesco Zappa~Nardelli, \emph{Verifying fence
  elimination optimisations}, Static Analysis (SAS), LNCS, vol. 6887, Springer,
  2011, pp.~146--162.

\end{thebibliography}
}

\iffull
\newpage
\appendix

\fi

\end{document}